\documentclass[aps,pre,twocolumn,10pt,superscriptaddress,nofootinbib,balancelastpage]{revtex4-2}
\usepackage[latin1]{inputenc}
\usepackage{amsmath,amsfonts,amssymb}
\usepackage{mathtools}
\usepackage{xcolor} 
\usepackage{seqsplit}
\usepackage{siunitx}
\usepackage{hyperref}\usepackage[capitalise]{cleveref}

\DeclareMathOperator{\Tr}{tr}       % Trace operator
\DeclareMathOperator{\diag}{diag}   % Diagonal operator
\newcommand{\norm}[1]{\lVert#1\rVert}%
\newcommand{\ZT}[1]{\textquotedblleft#1\textquotedblright}%

\allowdisplaybreaks

\begin{document}
\title{Inertial dynamics of an active Brownian particle}

\author{Jonas Mayer Martins}
\affiliation{Institute of Theoretical Physics, Center for Soft Nanoscience, University of M\"unster, 48149 M\"unster, Germany}

\author{Raphael Wittkowski}
\email[Corresponding author: ]{raphael.wittkowski@uni-muenster.de}
\affiliation{Institute of Theoretical Physics, Center for Soft Nanoscience, University of M\"unster, 48149 M\"unster, Germany}

\begin{abstract}
Active Brownian motion commonly assumes spherical overdamped particles. However, self-propelled particles are often neither symmetric nor overdamped yet underlie random fluctuations from their surroundings. Active Brownian motion has already been generalized to include asymmetric particles. Separately, recent findings have shown the importance of inertial effects for particles of macroscopic size or in low-friction environments. We aim to consolidate the previous findings into the general description of a self-propelled asymmetric particle with inertia. We derive the Langevin equation of such a particle as well as the corresponding Fokker-Planck equation. Furthermore, a formula is presented that allows reconstructing the hydrodynamic resistance matrix of the particle by measuring its trajectory. Numerical solutions of the Langevin equation show that, independently of the particle's shape, the noise-free trajectory at zero temperature starts with an inertial transition phase and converges to a circular helix. We discuss this universal convergence with respect to the helical motion that many microorganisms exhibit.
\end{abstract}
\maketitle

\section{Introduction}
The most basic model of Brownian motion assumes passive point particles without inertia. A first mathematical description was given by Einstein in 1905, which supported the then debated theory that matter is made of atoms \cite{Einstein1905c}.
There are three important generalizations to this basic model.

First, self-propelled particles, also referred to as active matter, generalize passive particles \cite{BechingerdLLRVV2016}. The study of active matter is highly relevant for biology, chemistry, and physics \cite{RomanczukBELSG2012, BechingerdLLRVV2016}. Active Brownian motion has proven a useful model for the nonequilibrium dynamics of self-propelled particles on the mesoscopic scale, such as bacteria \cite{LaugaP2009} and Janus particles \cite{KurzthalerDAFPMB2018}. Active colloids show various forms of emergent collective behavior \cite{ZoettlS2016}. The self-organization of light-driven particles is of particular interest for the development of intelligent matter \cite{AryaFKLS2019, DenzJRHJW2021}. On the mesoscopic scale, at low Reynolds numbers, inertial effects are negligible compared with the viscous friction that arises from the surrounding fluid. When inertia can be neglected, the particle's motion is called overdamped.

Second, inertia becomes relevant for macroscopic particles, such as vibrobots \cite{DebalisBGDVLBBK2018, ScholzJLL2018}, squid utilizing jet propulsion \cite{BartolKST2009}, or cars \cite{OroszWS2010}, and for motion in fluids of low viscosity, such as insect flight \cite{MukundarajanBKP2016}. The relevant time scale lies in between the Fokker-Planck scale and the Brownian scale \cite{Dhont1996} so that the influence of the solvent may well be modeled as a stochastic force, yet inertia cannot be neglected. Brownian motion that includes inertia is also called inertial Brownian motion \cite{Stolovitzky1998}, underdamped Brownian motion \cite{BodrovaCCSSM2016}, or Langevin motion \cite{Loewen2020}.
Some single-particle and collective effects of inertia on the dynamics of active particles have, for instance, been described in Ref.\ \cite{Loewen2020}.
Recently, the model of inertial Brownian motion has been extended to time-dependent parameters \cite{SprengerJIL2021}. Inertia has been shown to play a significant role, even at long times, in underdamped Brownian motion with time-dependent temperature \cite{BodrovaCCSSM2016}.

The third generalization concerns the particle's shape.
A colloidal particle's shape may be classified as isotropic, i.e., spherical, or anisotropic\footnote{\ZT{(An)isometric} is a more precise albeit less common term than \ZT{(an)isotropic} to describe the shape of a particle \cite{StenhammarWMC2015}.} \cite{WittkowskiL2011}.
The latter comprises uniaxial particles, which have an axis of rotational symmetry, like rods, dumbbells, or cones, as well as biaxial particles with discrete rotational symmetry or no symmetry at all, like oligomers and chiral particles \cite{WittkowskiL2011}.

A great part of the research on Brownian motion assumes microswimmers with symmetric shapes, such as spheres and ellipsoids \cite{tenHagenvTL2011, GMS2020} or rod-like particles \cite{Dhont1996, BaerGHP2020}.
Yet real particles are seldom highly symmetric. The description of Brownian motion has therefore been generalized to and experimentally verified for asymmetric particles in two spatial dimensions \cite{KuemmeltHWBEVLB2013, KuemmeltHWTBEVLB2014, ChakrabartyKWSSW2014}.
For the overdamped case, Ref.\ \cite{WittkowskiL2012} describes biaxial swimmers, whose noise-free trajectory at zero temperature in the absence of exterior forces is shown to be a circular helix. 

In summary, overdamped Brownian motion for arbitrarily shaped active particles on the one hand and inertial Brownian motion for simpler active particles on the other hand have been described in the literature. However, there has been no complete description so far that merges these two generalizations.

In this article, we study the Brownian motion of a self-propelled particle of arbitrary shape with inertia, consolidating the previous findings mentioned above into a more general model.
We provide the Langevin equation for such a particle. Our simulations of the particle's motion based on this Langevin equation show that, for every particle shape, the noise-free trajectory is a circular helix preceded by an initial transition phase due to inertia.
Furthermore, we derive the inertial Fokker-Planck equation for such a particle and present a formula that allows reconstructing the particle's shape-dependent hydrodynamic resistance matrix from its trajectory. This formula is analogous to the one presented in Ref.\ \cite{KraftWtHEPL2013} for overdamped Brownian motion.
Additionally, we consider the special cases of the Langevin equation that correspond to a particle in vacuum and to a particle whose motion is restricted to two spatial dimensions. 

This article is organized as follows: The next section presents the inertial Langevin equation, and its special cases for vacuum and for two spatial dimensions.
The inertial Fokker-Planck equation is derived in section \ref{sec:inertial_fokker-planck_equation}.
Section \ref{sec:numerical_methods} shows how we nondimensionalized and discretized the Langevin equation for performing simulations. It also provides details on the technicalities of the numerical solution.
Section \ref{sec:hydrodynamic_friction_tensor} gives the formula for the reconstruction of the hydrodynamic resistance matrix and tests the formula by applying it to numerical solutions of the Langevin equation.
Simulations using the Langevin equation for asymmetric particles and approaches to analytical solutions of the Langevin equation are discussed in section \ref{sec:simulation_and_analytical_solution}.
Finally, a summary of the results, concluding remarks, and an outlook on further research are given in section \ref{sec:conclusion}.

\section{\label{sec:inertial_Langevin_equation}Inertial Langevin equation}
The Langevin equation is a stochastic differential equation that describes the trajectory of a particle undergoing Brownian motion. We present in this section the Langevin equation for a particle with inertia and consider several special cases of it. Subsequently, we derive the inertial Fokker-Planck equation and the drift term that ensures thermodynamic consistency.

It is convenient to summarize forces and torques in a single vector, $\mathbf{K}_{\dots} = (\mathbf{F}_{\dots}, \mathbf{T}_{\dots})^\mathrm{T}$, which will be referred to as generalized force.
The inertial Langevin equation of an asymmetric self-propelled particle arises from the balance of forces $\mathbf{F}$ and torques $\mathbf{T}$ that act on the particle in the laboratory frame,
\begin{equation}
    \mathbf{K}_\mathrm{tot} = \mathbf{K}_\mathrm{H} + \mathbf{K}_\mathrm{int} + \mathbf{K}_\mathrm{ext} + \mathbf{K}_\mathrm{Br}\,,
\label{eq:forces}
\end{equation}
namely the total generalized force $\mathbf{K}_\mathrm{tot}$ that is equal to the sum of the hydrodynamic friction $\mathbf{K}_\mathrm{H}$, an internal generalized force $\mathbf{K}_\mathrm{int}$ that models the self-propulsion, an external generalized force $\mathbf{K}_\mathrm{ext}$, and lastly the Brownian noise $\mathbf{K}_\mathrm{Br}$.
We will now introduce each term in Eq.\ \eqref{eq:forces}.

First, $\mathbf{K}_\mathrm{tot}$ is the total generalized force
\begin{equation}
    \mathbf{K}_\mathrm{tot}
    = \mathcal{M} \frac{\partial \mathfrak{v}}{\partial t}\,,
\label{eq:newton}
\end{equation}
where the generalized mass
\begin{equation}
    \mathcal{M} = \begin{pmatrix}
                    m \mathrm{I} & 0 \\
                    0       & \mathrm{J}
                  \end{pmatrix}
\end{equation}
combines the mass $m$ and the moment of inertia tensor $\mathrm{J}$ of the particle. The generalized velocity $\mathfrak{v} = (\mathbf{v}, \boldsymbol{\omega})^\mathrm{T}$
combines the translational velocity $\mathbf{v}$ and angular velocity $\boldsymbol{\omega}$ of the particle. We use $\mathrm{I}$ to denote the $3\times3$--dimensional identity matrix. Equation \ref{eq:newton} is an extension of Newton's second law to orientational degrees of freedom.

$\mathbf{F}_\mathrm{H}$ and $\mathbf{T}_\mathrm{H}$ are the hydrodynamic friction force and torque acting on the particle when it moves through the surrounding fluid, which at low Reynolds number depend linearly on $\mathbf{v}$ and $\boldsymbol{\omega}$ through \cite{HappelB1991}
\begin{equation}
	\mathbf{K}_\mathrm{H} = - \eta \mathcal{R}^{-1} \mathcal{H} \mathcal{R} \mathfrak{v}
\end{equation}
with the $6\times6$--dimensional block diagonal matrix $\mathcal{R} = \diag (\mathrm{R}, \mathrm{R})$ of the $3\times3$--dimensional rotation matrix $\mathrm{R}$, and the hydrodynamic resistance matrix
\begin{equation}
	\mathcal{H} =
    	\begin{pmatrix}
    		\mathrm{K}            & \mathrm{C}_\mathrm{S}^\mathrm{T} \\
    		\mathrm{C}_\mathrm{S} & \Omega_\mathrm{S}
    	\end{pmatrix} ,
\label{eq:H_tensor_definition}
\end{equation}
whose components, with respect to the particle-fixed coordinate system that has its origin at the reference point $S$, depend on shape and size of the particle \cite{VossW2018, VossJW2019}. The $3\times3$--dimensional matrices $\mathrm{K}$ and $\Omega_\mathrm{S}$ are symmetric and determine translational-translational and rotational-rotational coupling, respectively, whereas $\mathrm{C}_\mathrm{S}$ may be asymmetric and describes the hydrodynamic translational-rotational coupling. In particular, $\mathcal{H}$ is diagonal for orthotropic particles, i.e., particles with three pairwise orthogonal symmetry planes. A detailed description of the relation between the structure of the tensor and a particle's shape can be found in Ref.\ \cite{VossW2018}.

The short-time diffusion tensor $\mathcal{D}$ is inverse to the hydrodynamic resistance matrix,
\begin{equation}
	\mathcal{D} = \frac{1}{\beta \eta} \mathcal{R}^{-1} \mathcal{H}^{-1} \mathcal{R}\,,
\label{eq:diffusion_tensor_definition}
\end{equation}
where the thermodynamic beta is defined as $\beta = 1/(k_\mathrm{B} T)$ with the Boltzmann constant $k_\mathrm{B}$ and temperature $T$ of the fluid surrounding the particle, and $\eta$ means the viscosity of the fluid.

The rotation matrix $\mathrm{R}$ describes the particle's orientation and transforms the frame of reference from laboratory to body frame. We write the generalized position of the particle as the tuple $\mathfrak{x} = (\mathbf{x}, \mathrm{R})^\mathrm{T}$, where the vector $\mathbf{x}$ denotes the particle's position.
The Euler angles\footnote{There is a variety of conventions for Euler angles. We use the convention of Gray and Gubbins \cite{GrayG1984}, which generalizes the spherical coordinates $\theta$ and $\phi$ with a third angle $\chi$.}
$\boldsymbol{\varpi} = (\phi, \theta, \chi)^\mathrm{T}$, which characterize the particle's orientation equivalently, generate the rotation matrix $\mathrm{R}$ as a product of elementary rotation matrices through
\begin{equation}
	\mathrm{R}(\boldsymbol{\varpi}) = \mathrm{R}_3(\chi) \mathrm{R}_2(\theta) \mathrm{R}_3(\phi) \,,
\label{eq:rotation_matrix_definition}
\end{equation}
where $\mathrm{R}_i(\varphi)$ describes a clockwise rotation by the angle $\varphi$ about the $i$th Cartesian axis, looking down the axis, for $i=1,2,3$. The elementary matrices are
\begin{align}
	\mathrm{R}_1(\varphi) &=
	\begin{pmatrix*}[r]
		1 		& 0\quad\; 			& 0\quad\;	    \\
		0 		& \cos(\varphi) 	& \sin(\varphi) \\
		0		& -\sin(\varphi) 	& \cos(\varphi)
	\end{pmatrix*}, 
	\label{eq:elementary_rotation_matrices_definition_1} \\
	\mathrm{R}_2(\varphi) &=
	\begin{pmatrix*}[r]
		\cos(\varphi) 	& 0 & -\sin(\varphi)    \\
		0\quad\; 		& 1 & 0\quad\; 			\\
		\sin(\varphi)	& 0	& \cos(\varphi)
	\end{pmatrix*}, 
	\label{eq:elementary_rotation_matrices_definition_2} \\
	\mathrm{R}_3(\varphi) &=
	\begin{pmatrix*}[r]
		\cos(\varphi) 	& \sin(\varphi) & 0	\\
		-\sin(\varphi) 	& \cos(\varphi) & 0 \\
		0\quad\;	    & 0\quad\;		& 1
	\end{pmatrix*}
    \label{eq:elementary_rotation_matrices_definition_3}
\end{align}
and the rotation matrix is orthogonal so that $\mathrm{R}^{-1} = \mathrm{R}^\mathrm{T}$.

The internal generalized force $\mathbf{K}_\mathrm{int}$ describes an effective force and torque modeling the particle's active motion in the laboratory frame. They can be written in terms of the vectors $\mathbf{F}_0$ and $\mathbf{T}_0$ in the body frame as $\mathbf{F}_\mathrm{int} = \mathrm{R}^{-1} \mathbf{F}_0$ and $\mathbf{T}_\mathrm{int} = \mathrm{R}^{-1} \mathbf{T}_0$, i.e., $\mathbf{K}_\mathrm{int}=\mathcal{R}^{-1}\mathbf{K}_0$ with $\mathbf{K}_0=(\mathbf{F}_0,\mathbf{T}_0)^\mathrm{T}$.
It has been justified in Ref.\ \cite{tenHagenWTKBL2015} that the description of active motion through effective forces and torques is viable.

Next, $\mathbf{K}_\mathrm{ext}$ describes the external force and torque. This could, for example, be a gravitational or magnetic field acting on the particle. If $\mathbf{K}_\mathrm{ext}$ is conservative, it can be written as the negative gradient of a \mbox{potential $\Phi(\mathfrak{x})$}, i.e., $\mathbf{K}_\mathrm{ext} = -\nabla_\mathfrak{x} \Phi$, where the generalized gradient \mbox{$\nabla_{\mathfrak{x}} = (\nabla_{\mathbf{x}}, \nabla_\mathrm{R})^\mathrm{T}$} consists of the translational gradient $\nabla_{\mathbf{x}} = (\partial_{x_1}, \partial_{x_2}, \partial_{x_3})^\mathrm{T}$ and the rotational gradient \mbox{$\nabla_\mathrm{R} = \mathrm{i}\mathbf{L}$} with angular momentum operator $\mathbf{L}$. The rotational gradient is considered in more detail in section \ref{sec:inertial_fokker-planck_equation}.

Finally, the Brownian generalized force $\mathbf{K}_\mathrm{Br}$ models random collisions between the molecules of the fluid and the particle. This interaction can be characterized, in the body frame, by a stochastic force $\mathbf{f}_0(t)$ and \mbox{torque $\boldsymbol{\tau}_0(t)$} at \mbox{time $t$}. In the laboratory frame, this gives $\mathbf{K}_\mathrm{Br} = \mathcal{R}^{-1} \mathbf{k} + \mathbf{K}_\mathrm{D}$ with the generalized stochastic force $\mathbf{k} = (\mathbf{f}_0, \boldsymbol{\tau}_0)^\mathrm{T}$ in the body frame. The drift term $\mathbf{K}_\mathrm{D}$ ensures that the Langevin \mbox{equation \eqref{eq:langevin}} (see below) conforms with the Boltzmann distribution in thermal equilibrium \cite{LauL2007}. In section \ref{sec:inertial_fokker-planck_equation}, we show based on the Fokker-Planck equation that no drift term is necessary for inertial Brownian motion, i.e., $\mathbf{K}_\mathrm{D} = \mathbf{0}$, if the external generalized force $\mathbf{K}_\mathrm{ext}$ is conservative.

The characteristics of the noise $\mathbf{k}(t)$ still need to be specified. We assume that $\mathbf{k}(t)$ is Gaussian white noise with ensemble average
\begin{equation}
\langle \mathbf{k}(t) \rangle = \mathbf{0}
\label{eq:noise_mean}
\end{equation}
and auto-correlation
\begin{equation}
\langle \mathbf{k}(t_1) \otimes \mathbf{k}(t_2) \rangle = \sigma^2 \delta(t_1 - t_2)\,,
\label{eq:noise_variance}
\end{equation}
where $\sigma = \sqrt{(2 \eta / \beta) \mathcal{H}}$ is the standard deviation matrix\footnote{Since the hydrodynamic resistance matrix $\mathcal{H}$ is real and symmetric, it can be written as $\mathcal{H}=\mathrm{P}\diag(\lambda_1,\dotsc,\lambda_6)\mathrm{P}^{-1}$ with an orthogonal matrix $\mathrm{P}$ and the eigenvalues $\lambda_1,\dotsc,\lambda_6$ of $\mathcal{H}$. Thus, the square root of $\mathcal{H}$ is defined as $\sqrt{\mathcal{H}}=\mathrm{P}\diag(\sqrt{\lambda_1},\dotsc,\sqrt{\lambda_6})\mathrm{P}^{-1}$.}.
Larger eigenvalues of the standard deviation matrix $\sigma$ lead to stronger fluctuations.
The mean of the noise is zero because the fluctuating force and torque are not directed. The delta correlation $\delta(t_1-t_2)$ means that the correlation time is zero, so no nudge has memory of any nudges that happened before (Markov property). The description is therefore limited to times much larger than the time scale of the solvent molecules, which is typically around $10^{-14}\,\mathrm{s}$, since only then is the correlation time negligible \cite{Dhont1996}.
In contrast, noise with nonzero correlation time is called colored noise \cite{ElberMO2020}.

Inserting the definitions above into the balance of forces and torques \eqref{eq:forces}, we obtain the \emph{inertial Langevin equation} of an asymmetric self-propelled particle
\begin{equation}
\begin{split}
    \frac{\partial \mathfrak{v}}{\partial t} &= \mathcal{M}^{-1} \big(
            - \eta \mathcal{R}^{-1} \mathcal{H} \mathcal{R} \mathfrak{v}
            + \mathcal{R}^{-1} \mathbf{K}_0 + \mathbf{K}_\mathrm{ext} +  \mathbf{K}_\mathrm{D}
        \big) \\
        &\quad\,+ \mathcal{M}^{-1} \mathcal{R}^{-1} \mathbf{k} \,.
\raisetag{3.85ex}
\label{eq:langevin}
\end{split}
\end{equation}
This Langevin equation constitutes the first main result of this article. 

\subsection{Overdamped particles}
At times much larger than the momentum relaxation time $\tau_\mathrm{p}$ of the particle, the ensemble average $\langle \mathfrak{p}(t) \otimes \mathfrak{p}(t) \rangle$ with the generalized momentum $\mathfrak{p} = \mathcal{M}\mathfrak{v}$ and the dyadic product $\otimes$ is in equilibrium and inertia can be neglected, i.e., $\mathbf{K}_\mathrm{tot}=\mathbf{0}$ in the balance of forces and torques \eqref{eq:forces}.
The auto-correlation matrix \cite{Dhont1996}
\begin{equation}
\begin{split}
    & \big\langle
        (\mathbf{p}(t) - \langle\mathbf{p}(t)\rangle)
        \otimes
        (\mathbf{p}(t) - \langle\mathbf{p}(t)\rangle)
    \big\rangle \\
    & = \mathrm{I} \frac{m}{\beta}
    \Big(1 - \exp{\!\Big(- 2\frac{\gamma}{m} t\Big)} \big)
\end{split}
\end{equation}
determines the momentum relaxation time $\tau_\mathrm{p} = m/\gamma$ for a spherical particle of mass $m$ and friction coefficient $\gamma = 6 \pi \eta r$ with the particle's radius $r$. In the general case of an asymmetric particle, we have found no closed expression for the momentum relaxation time.
The overdamped drift term is
$\mathbf{K}_\mathrm{D} = (\beta \mathcal{D}(\mathbf{\mathfrak{x}}))^{-1} \mathbf{\nabla}_{\mathbf{\mathfrak{x}}} \mathcal{D}(\mathbf{\mathfrak{x}})$ \cite{WittkowskiL2012}.
Hence, the overdamped Langevin equation for a biaxial particle described in Ref.\ \cite{WittkowskiL2012},
\begin{equation}
	\mathfrak{v} = \beta \mathcal{D}
	\big(\mathcal{R}^{-1} \mathbf{K}_{0} + \mathbf{K}_\mathrm{ext}
	+ \mathbf{K}_\mathrm{D} \big) + \beta \mathcal{D} \mathcal{R}^{-1} \mathbf{k}\,,
    \label{langevin_overdamped}
\end{equation}
is a limiting case of the inertial Langevin equation \eqref{eq:langevin}.

\subsection{\label{sec:langevin_in_vacuum}Particles in vacuum}
Some propulsion methods of active particles, such as light-driven particles, work even without a surrounding medium \cite{DenzJRHJW2021}.
In vacuum, there is no hydrodynamic friction, i.e., $\mathbf{K}_\mathrm{H} = \mathbf{0}$, and there is no stochastic force either without a surrounding medium, i.e., $\mathbf{K}_\mathrm{Br}=\mathbf{0}$. The Langevin equation thus simplifies to the deterministic equation of motion
\begin{equation}
    \frac{\partial \mathfrak{v}}{\partial t} =
        \mathcal{M}^{-1}
        (\mathbf{K}_\mathrm{int} + \mathbf{K}_\mathrm{ext}) \,.
\end{equation}
This is just what we would expect for a self-propelled particle in vacuum.

In this equation, the translational velocity $\mathbf{v}$ and angular velocity $\boldsymbol{\omega}$ are no longer directly coupled:
\begin{align}
    \frac{\partial \mathbf{v}}{\partial t} &=
        m^{-1} (\mathrm{R}^{-1} \mathbf{F}_0 + \mathbf{F}_\mathrm{ext} )\,, \\
    \frac{\partial \boldsymbol{\omega}}{\partial t} &=
        \mathrm{J}^{-1} (\mathrm{R}^{-1} \mathbf{T}_0 + \mathbf{T}_\mathrm{ext} )\,.
\end{align}
The forces $\mathbf{F}_0$ and $\mathbf{F}_\mathrm{ext}$ and torques $\mathbf{T}_0$ and $\mathbf{T}_\mathrm{ext}$, however, can still depend on the position $\mathbf{x}$ and \mbox{orientation $\mathrm{R}$}.

\subsection{\label{sec:langevin_in_2D}Particles in two spatial dimensions}
The Langevin equation \eqref{eq:langevin} in its full generality can only be solved numerically. To obtain analytical solutions, we must resort to simplifications. One possible constraint is to reduce the motion to two spatial dimensions.

Analogously to Ref.\ \cite{WittkowskiL2012} for the overdamped case, we set $x_3 = 0$ to constrain translation to the $x_1$,$x_2$-plane, and restrict rotation to the azimuthal angle $\phi$ by choosing $\theta = \pi/2$ and $\chi = 0$.
Next, we set $\mathcal{R}^{-1} \mathbf{K}_0 = (\mathbf{F}_\mathrm{A},0,0,0,T_\mathrm{A})^\mathrm{T}$ with internal force $\mathbf{F}_\mathrm{A}$ and internal torque $T_\mathrm{A}$. Similarly, we set $\mathbf{K}_\mathrm{ext} = (\mathbf{F}_\mathrm{ext},0,0,0,T_\mathrm{ext})^\mathrm{T}$. The stochastic force $\mathbf{f}$ and torque $\tau$ define
$\mathbf{k} = (0, f_\bot, f_\parallel, -\tau, 0, 0)^\mathrm{T}$
and
$\mathcal{R}^{-1} \mathbf{k} = (\mathbf{f},0,0,0,\tau)^\mathrm{T}$.
Note that the stochastic force can be decomposed as 
$\mathbf{f} = f_\parallel \mathbf{u}_\parallel + f_\bot \mathbf{u}_\bot$ with regard to the particle's orientation
vector in the plane $\mathbf{u}_\parallel(\phi) = (\cos(\phi), \sin(\phi))^\mathrm{T}$ and its orthogonal counterpart $\mathbf{u}_\bot(\phi) = (-\sin(\phi), \cos(\phi))^\mathrm{T}$. The internal force can be decomposed similarly as $\mathbf{F}_\mathrm{A} = F_{\mathrm{A},\parallel} \mathbf{u}_\parallel + F_{\mathrm{A},\bot} \mathbf{u}_\bot$.

The two orientation vectors $\mathbf{u}_\parallel$ and $\mathbf{u}_\bot$ give rise to the $2\times2$--dimensional rotation matrix
\begin{equation}
	\mathrm{\widetilde{R}}(\phi) = \begin{pmatrix*}[r]
  		\cos(\phi) & -\sin(\phi) \\
  		\sin(\phi) &  \cos(\phi)
	\end{pmatrix*}
	=
	\big(
	    \mathbf{u}_\parallel(\phi), \mathbf{u}_\bot(\phi)
	\big) ,
\end{equation}
whose inclusion in three spatial dimensions is the matrix
\begin{equation}
	\mathrm{R}(\phi) = \begin{pmatrix*}[r]
  		0 & -\sin(\phi) & \cos(\phi) \\
  		0 &  \cos(\phi) & \sin(\phi) \\
 		-1& 0 \quad\;     &  0 \quad\;
	\end{pmatrix*}.
\label{rotation_matrix_restricted}
\end{equation}
For this part of the section, we write the block diagonal matrix $\mathcal{R} = \diag (\mathrm{R}, \mathrm{R})$ with the rotation matrix from Eq.\ \ref{rotation_matrix_restricted}.

With these definitions taken care of, the question remains which constraints must be imposed on $\mathcal{H}$ to ensure that motion stays in two spatial dimensions.
The definitions of the restricted rotation matrix \eqref{rotation_matrix_restricted} and orientation vectors $\mathbf{u}_\parallel$ and $\mathbf{u}_\perp$ let us write the hydrodynamic friction as
\begin{equation}
\begin{split}
  \mathcal{R} \mathcal{H} \mathcal{R}^{-1} \mathfrak{v}
  &=
  \mathcal{R} \mathcal{H}
  \begin{pmatrix} 0 \\ \mathbf{u}_\bot v_1 + \mathbf{u}_\parallel v_2 \\
                - \omega_3  \\ 0 \\ 0 \end{pmatrix} \\
  &=
  \mathcal{R}
  \begin{pmatrix}
    H_{12} v_\perp + H_{13} v_\parallel - H_{14} \omega \\
    H_{22} v_\perp + H_{23} v_\parallel - H_{24} \omega \\
    H_{32} v_\perp + H_{33} v_\parallel - H_{34} \omega \\
    H_{42} v_\perp + H_{43} v_\parallel - H_{44} \omega \\
    H_{52} v_\perp + H_{53} v_\parallel - H_{54} \omega \\
    H_{62} v_\perp + H_{63} v_\parallel - H_{64} \omega \\
  \end{pmatrix},
\end{split}
\end{equation}
where we abbreviate $v_\perp(\phi) = -\sin(\phi)\, v_1 + \cos(\phi)\, v_2$ and $v_\parallel(\phi) = \cos(\phi)\, v_1 + \sin(\phi)\, v_2$, and where $v_1$, $v_2$, and $\omega_3$ are the respective components of $\mathbf{v}$ and $\boldsymbol{\omega}$.
It can be easily seen that all elements of $\mathcal{H}$ must be zero, except for $\widetilde{\mathcal{H}} = (H_{ij})_{i,j = 2,3,4}$, since only then is the hydrodynamic friction,
\begin{equation}
\begin{split}
  \dotsm &=
  \begin{pmatrix}
    R & 0 \\
    0 & R
  \end{pmatrix}
  \begin{pmatrix}
    0 \\
    H_{22} v_\perp + H_{23} v_\parallel - H_{24} \omega \\
    H_{32} v_\perp + H_{33} v_\parallel - H_{34} \omega \\
    H_{42} v_\perp + H_{43} v_\parallel - H_{44} \omega \\
    0 \\
    0 \\
  \end{pmatrix} \\
  &=
  \begin{pmatrix}
    \quad (H_{22} v_\perp + H_{23} v_\parallel - H_{24} \omega)\mathbf{u}_\bot \\
    + (H_{32} v_\perp + H_{33} v_\parallel - H_{34} \omega) \mathbf{u}_\parallel \\
    0 \\
    0 \\
    0 \\
    - (H_{42} v_\perp + H_{43} v_\parallel - H_{44} \omega)
  \end{pmatrix},
\end{split}
\label{eq:hydrodynamic_friction_restricted}
\end{equation}
a vector that is zero in all elements that would induce movement outside the $x_1$,$x_2$-plane and \mbox{$\omega_3$ axis}. If the hydrodynamic friction had a nonzero element in the direction of $v_3$, $\omega_1$, or $\omega_2$, the particle would leave the two-dimensional setting.

Since $\mathcal{H}$ is symmetric, i.e., $H_{ij} = H_{ji}$, it is well defined to set
\begin{equation}
	\begin{pmatrix}
		H_3 & H_2 & H_\mathrm{C}^\bot \\
		H_2 & H_1 & H_\mathrm{C}^\parallel \\
		H_\mathrm{C}^\bot & H_\mathrm{C}^\parallel & H_\mathrm{R} \\
	\end{pmatrix}
	=
	\begin{pmatrix}
		H_{22} & H_{23} & H_{24} \\
		H_{32} & H_{33} & H_{34} \\
		H_{42} & H_{43} & H_{44} \\
	\end{pmatrix}.
\end{equation}
For the rest of this section, we redefine $\mathbf{v} = (v_1, v_2)^\mathrm{T}$ and $\omega = \omega_3$ to simplify the notation.
We insert the result for the hydrodynamic friction from Eq.\ \eqref{eq:hydrodynamic_friction_restricted} into the restricted Langevin equation \eqref{eq:langevin} to obtain the two equations
\begin{align}
\begin{split}
  m \smash{\frac{\partial \mathbf{v}}{\partial t}}
  &= - \eta \big(
    (H_3 v_\perp +
     H_2 v_\parallel -
     H_\mathrm{C}^\bot \omega) \mathbf{u}_\bot \\
  &\qquad\quad + (H_2 v_\perp + H_1 v_\parallel - H_\mathrm{C}^\parallel \omega) \mathbf{u}_\parallel \big)  \\
  &\quad\, + \mathbf{F}_\mathrm{A} + \mathbf{F}_\mathrm{ext} + \mathbf{f} \,,
\label{langevin_restriced_velocity_expanded}
\end{split} \\
\begin{split}
  \widetilde{J} \frac{\partial \omega}{\partial t}
  &= \eta \big(H_\mathrm{C}^\bot v_\perp + H_\mathrm{C}^\parallel v_\parallel - H_\mathrm{R} \omega \big) \\
  &\quad + T_\mathrm{A} + T_\mathrm{ext} + \tau \,.
\end{split}
\end{align}
The coefficients $H_\mathrm{C}^\bot$ and $H_\mathrm{C}^\parallel$, combined in the coupling vector $\mathbf{H}_\mathrm{C} = H_\mathrm{C}^\parallel \mathbf{u}_\parallel + H_\mathrm{C}^\bot \mathbf{u}_\bot$, couple translational and rotational motion.
The submatrix
\begin{equation}
	\mathrm{H} = \begin{pmatrix}
		H_1 & H_2 \\
		H_2 & H_3
		\end{pmatrix}
\end{equation}
accounts for the translational-translational coupling, whereas $H_\mathrm{R}$ causes the rotational-rotational coupling.
Since the rotation is restricted to the $\omega_3$-axis, the relevant part of inertia $\mathrm{J}$ is just the element $\widetilde{J} = J_{33}$.

Finally, we reinsert the definitions of $v_\perp$ and $v_\parallel$ into the translational part \eqref{langevin_restriced_velocity_expanded} to get the coupled two-dimensional inertial Langevin equations
\begin{align}
	m \frac{\partial \mathbf{v}}{\partial t} &=
	    - \eta \big( \widetilde {\mathrm{R}} \mathrm{H} \widetilde{\mathrm{R}}^{-1} \mathbf{v}
        - \mathbf{H}_\mathrm{C} \omega \big)+ \mathbf{F}_\mathrm{A} + \mathbf{F}_\mathrm{ext} + \mathbf{f} \,, \label{eq:langevin_translational_2D} \\
        \widetilde{J} \frac{\partial \omega}{\partial t}
    &= \eta \big(\mathbf{H}_\mathrm{C} \widetilde{\mathrm{R}}^{-1} \mathbf{v} - H_\mathrm{R} \omega \big)
        + T_\mathrm{A} + T_\mathrm{ext} + \tau \,, \label{eq:langevin_rotational_2D}
\end{align}
which bear remarkable resemblance to their three-dimensional counterpart \eqref{eq:langevin}.

\section{\label{sec:inertial_fokker-planck_equation}Inertial Fokker-Planck equation} 
For our derivation of the inertial Fokker-Planck equation, following the approach of Ref.\ \cite{ElberMO2020},
we are interested in the time derivative of the probability density function
\begin{equation}
	P(\mathfrak{x}, \mathfrak{v}, t) = \langle \delta(\mathfrak{x} - \mathfrak{x}(t)) \delta(\mathfrak{v} - \mathfrak{v}(t)) \rangle \,,
\label{eq:pdf}
\end{equation}
where $\langle \cdot \rangle$ denotes the average over all noise realizations.
We define the infinitesimal time evolution of position $\mathbf{x}$, orientation $\mathrm{R}$, and generalized velocity $\mathfrak{v}$ as
\begin{align}
  \mathbf{x}(t + \mathrm{d}t)   &= \mathbf{x}(t)    + \mathrm{d}\mathbf{x}(t) \,,  \\
  \mathrm{R}(t + \mathrm{d}t)   &= \mathrm{R}(t)    + \mathrm{d}\mathrm{R}(t)   \,,  \\
  \mathfrak{v}(t + \mathrm{d}t) &= \mathfrak{v}(t)  + \mathrm{d}\mathfrak{v}(t) \,.
\end{align}
A Taylor expansion around $\mathbf{x} - \mathbf{x}(t)$ etc. up to less than second order in $\mathrm{d}t$ (which means up to $(\mathrm{d}\mathfrak{v})^2$ since $\mathrm{d}\mathfrak{v} \sim \sqrt{\mathrm{d}t}$, and up to $\mathrm{d}\mathfrak{x}$) gives
\begin{align}
    \begin{split}
 	\delta(\mathbf{x} - \mathbf{x}(t) - \mathrm{d}\mathbf{x}(t))
 	&= \big(1 - \mathrm{d}\mathbf{x}^\mathrm{T}(t) \nabla_\mathbf{x} \big) \delta(\mathbf{x} - \mathbf{x}(t)) \,,
 	\label{eq:fokker-planck_derivation_expansions_position}
  	\end{split} \raisetag{6ex}\\
  	\delta(\mathrm{R} - \mathrm{R}(t) - \mathrm{d}\mathrm{R}(t))
  	&= \big(1 - \mathrm{d}t \,\boldsymbol{\omega}^\mathrm{T}(t) \, \nabla_\mathrm{R} \big) \delta(\mathrm{R} - \mathrm{R}(t)) \,,
  	\label{eq:fokker-planck_derivation_expansions_rotation} \\
  	\begin{split}
  	\delta(\mathfrak{v} - \mathfrak{v}(t) - \mathrm{d}\mathfrak{v}(t))
  	&= \bigg(1 - \mathrm{d}\mathfrak{v}^\mathrm{T} \nabla_\mathfrak{v}
	+ \frac{1}{2} \mathrm{d}\mathfrak{v}^\mathrm{T}(t) \nabla_{\mathfrak{v}} \nabla_{\mathfrak{v}}^\mathrm{T} \mathrm{d} \mathfrak{v}(t) \bigg) \\ & \quad\;\, \cdot \delta(\mathfrak{v} - \mathfrak{v}(t)) \,,
	\raisetag{3ex}
	\label{eq:fokker-planck_derivation_expansions_velocity}
    \end{split}
\end{align}
where $ \nabla_{\mathfrak{v}} \nabla_{\mathfrak{v}}^\mathrm{T} \delta(\mathfrak{v} - \mathfrak{v}(t))$ is the Hessian matrix of $\delta(\mathfrak{v}-\mathfrak{v}(t))$, $\nabla_\mathrm{R} = \mathrm{i}\mathbf{L}$ the rotational gradient operator, and $\mathbf{L}$ the angular momentum operator. The expansion for the rotational part is not immediately obvious. We will briefly present the underlying argument:

The distributions $\delta(\mathrm{R} - \mathrm{R}(t) - \mathrm{d}\mathrm{R}(t))$ and $\delta(\boldsymbol{\varpi} - \boldsymbol{\varpi}(t) - \mathrm{d}\boldsymbol{\varpi}(t))$ are obviously equivalent. The latter may be more easily expanded than the matrix-valued distribution as
\begin{equation}
	\delta(\boldsymbol{\varpi} - \boldsymbol{\varpi}(t) - \mathrm{d}\boldsymbol{\varpi}(t))
	= \big(1 - \mathrm{d}\boldsymbol{\varpi}^\mathrm{T}(t) \nabla_{\boldsymbol{\varpi}} \big) \delta(\boldsymbol{\varpi} - \boldsymbol{\varpi}(t))\,.
\end{equation}
We consider the matrix
\begin{equation}
	\mathrm{M}(\boldsymbol{\varpi}) =
	\begin{pmatrix*}[r]
		0 	& -\sin(\phi)	& \cos(\phi) \sin(\theta) \\
		0 	&  \cos(\phi)	& \sin(\phi) \sin(\theta) \\
		1 	& 0					& \cos(\theta)
	\end{pmatrix*},
\label{eq:fokker-planck_derivation_M_euler}
\end{equation}
introduced in Ref.\ \cite{Schutte1976}.
Its inverse is
\begin{equation}
	\mathrm{M}^{-1}(\boldsymbol{\varpi}) =
	\begin{pmatrix*}[l]
		- \cos(\phi) \cot(\theta)	        & - \sin(\phi) \cot(\theta)	            & 1 \\
		- \sin(\phi)	                    & \phantom{+} \cos(\phi)	            & 0 \\
	  	\phantom{+} \cos(\phi) \csc(\theta)	& \phantom{+} \sin(\phi) \csc(\theta)	& 0
	\end{pmatrix*}.
\end{equation}

The matrix $\mathrm{M}$ is important because it connects angular velocity $\boldsymbol{\omega}$ and Euler angles $\boldsymbol{\varpi}$ through \cite{Schutte1976}
\begin{equation}
	\boldsymbol{\omega} = \mathrm{M}(\boldsymbol{\varpi}) \frac{\mathrm{d}\boldsymbol{\varpi}}{\mathrm{d}t}
\label{eq:fokker-planck_derivation_euler_to_angular}
\end{equation}
and generates the rotational gradient operator
\begin{equation}
	\nabla_\mathrm{R} =
	\mathrm{i} \mathbf{L} =
	(\mathrm{M}^{-1} (\boldsymbol{\varpi}))^\mathrm{T} \nabla_{\boldsymbol{\varpi}} \,,
\label{eq:fokker-planck_derivation_iL_euler}
\end{equation}
which in combination leads to
\begin{equation}
	\mathrm{d}t \, \boldsymbol{\omega}^\mathrm{T} \, \mathrm{i} \mathbf{L} 
	= \mathrm{d}\boldsymbol{\varpi}^\mathrm{T} \nabla_{\boldsymbol{\varpi}} \,.
\label{eq:fokker-planck_derivation_rotational_gradient_proof}
\end{equation}

Returning to the original distribution $\delta(\mathrm{R} - \mathrm{R}(t))$, we get
\begin{equation}
	\big(1 - \mathrm{d}\boldsymbol{\varpi}^\mathrm{T} \nabla_{\boldsymbol{\varpi}} \big) \delta(\boldsymbol{\varpi} - \boldsymbol{\varpi}(t))
	= \big(1 - \mathrm{d}t \,\boldsymbol{\omega}^\mathrm{T} \, \mathrm{i} \mathbf{L} \big) \delta(\mathrm{R} - \mathrm{R}(t))\,,
\end{equation}
which leads to the proposed orientational expansion \eqref{eq:fokker-planck_derivation_expansions_rotation} and concludes the argument.

\begin{widetext}
With the expansions \eqref{eq:fokker-planck_derivation_expansions_position}-\eqref{eq:fokker-planck_derivation_expansions_velocity}, the differential of $P$ can be written as
\begin{equation}
\begin{split}
	\mathrm{d}P &= P(\mathbf{x}, \mathrm{R}, \mathfrak{v}, t + \mathrm{d}t)
    - P(\mathbf{x}, \mathrm{R}, \mathfrak{v}, t) \\
   	&= \langle
        \delta(\mathbf{x}   - \mathbf{x}(t + \mathrm{d}t))
        \delta(\mathrm{R}   - \mathrm{R}(t + \mathrm{d}t))
        \delta(\mathfrak{v} - \mathfrak{v}(t + \mathrm{d}t))
      - \delta(\mathbf{x}   - \mathbf{x}(t))
        \delta(\mathrm{R}   - \mathrm{R}(t))
        \delta(\mathfrak{v} - \mathfrak{v}(t)) \rangle \\
    &= \bigg\langle \big(1 - \mathrm{d}\mathbf{x}^\mathrm{T} \nabla_\mathbf{x} \big)
   \big(1 - \mathrm{d}t \,\boldsymbol{\omega}^\mathrm{T} \nabla_\mathrm{R} \big)
    \bigg(1 - \mathrm{d}\mathfrak{v}^\mathrm{T}  \nabla_\mathfrak{v}
    + \frac{1}{2} \mathrm{d}\mathfrak{v}^\mathrm{T} \nabla_{\mathfrak{v}} \nabla_{\mathfrak{v}}^\mathrm{T} \mathrm{d}\mathfrak{v} \bigg) 
    \delta(\mathbf{x}   - \mathbf{x}(t))
    \delta(\mathrm{R}   - \mathrm{R}(t))
    \delta(\mathfrak{v} - \mathfrak{v}(t)) \\
    &\qquad -
    \delta(\mathbf{x}   - \mathbf{x}(t))
    \delta(\mathrm{R}   - \mathrm{R}(t))
    \delta(\mathfrak{v} - \mathfrak{v}(t)) \bigg\rangle \,.
\end{split}
\end{equation}
Expanding the terms in parentheses in the last expression, again neglecting terms of higher order, we find
\begin{equation}
\begin{split}
	\mathrm{d}P = \bigg\langle
    \bigg(
      - \mathrm{d}\mathbf{x}^\mathrm{T} \nabla_\mathbf{x}
      - \mathrm{d}t \,\boldsymbol{\omega}^\mathrm{T} \nabla_\mathrm{R}
      - \mathrm{d}\mathfrak{v}^\mathrm{T} \nabla_\mathfrak{v}
      + \frac{1}{2} \mathrm{d}\mathfrak{v}^\mathrm{T}  \nabla_{\mathfrak{v}} \nabla_{\mathfrak{v}}^\mathrm{T} \mathrm{d}\mathfrak{v}
    \bigg)
    \delta(\mathbf{x} - \mathbf{x}(t))
    \delta(\mathrm{R} - \mathrm{R}(t))
    \delta(\mathfrak{v} - \mathfrak{v}(t)) \bigg\rangle \,.
\end{split}
\end{equation}
We know that $\mathrm{d}\mathbf{x} = \mathbf{v} \mathrm{d}t$, and the Langevin equation \eqref{eq:langevin} provides an expression for $\mathrm{d}\mathfrak{v}$, which results in
\begin{equation}
\begin{split}
	\mathrm{d}P &= \bigg\langle \bigg(
      - \mathbf{v}^\mathrm{T} \mathrm{d}t \nabla_\mathbf{x}
      - \mathrm{d}t \,\boldsymbol{\omega}^\mathrm{T}  \nabla_\mathrm{R}
      - \nabla_\mathfrak{v}^\mathrm{T} \mathcal{M}^{-1}
      \big(-\eta \mathcal{R}^{-1} \mathcal{H}
      	\mathcal{R} \mathfrak{v} + \mathcal{R}^{-1} \mathbf{K}_0
      	+ \mathbf{K}_\mathrm{ext} +  \mathbf{K}_\mathrm{D}
      \big) \mathrm{d}t 
      - \nabla_\mathfrak{v}^\mathrm{T} \mathcal{M}^{-1} \mathcal{R}^{-1} \mathrm{d}\mathbf{B} \\
      &\qquad + \frac{1}{2} (\mathbf{N}^\mathrm{T} \sigma \mathcal{R} \mathcal{M}^{-1}) \nabla_{\mathfrak{v}} \nabla_{\mathfrak{v}}^\mathrm{T} (\mathcal{M}^{-1}  \mathcal{R}^{-1} \sigma \mathbf{N})  \mathrm{d}t
   \bigg) \delta(\mathbf{x} - \mathbf{x}(t))
   \delta(\mathrm{R} - \mathrm{R}(t))
   \delta(\mathfrak{v} - \mathfrak{v}(t)) \bigg\rangle \,,
\end{split}
\label{eq:fokker-planck_derivation_langevin_inserted}
\end{equation}
where $\mathbf{N}$ means a 6-dimensional vector with standard normal distribution $\mathcal{N}(0,1)$ for each element.
The particle state $(\mathfrak{x},\mathfrak{v})^\mathrm{T}(t)$ is only affected by noise from the past, $\mathrm{d}\mathbf{B}(t')$ for $t' < t$, so that the current particle state is uncorrelated with the noise in the present. This causality principle implies 
for the stochastic noise
$\mathrm{d}\mathbf{B}(t) = \sqrt{\mathrm{d}t \mathcal{H} 2 \eta / \beta}\, \mathbf{N}$
that
\begin{equation}
\begin{split}
  \langle \mathrm{d}\mathbf{B}(t) \delta(\mathbf{x} - \mathbf{x}(t))
  \delta(\mathrm{R} - \mathrm{R}(t))
  \delta(\mathfrak{v} - \mathfrak{v}(t)) \rangle
  = \langle \mathrm{d}\mathbf{B}(t) \rangle
  \langle \delta(\mathbf{x} - \mathbf{x}(t))
  \delta(\mathrm{R} - \mathrm{R}(t))
  \delta(\mathfrak{v} - \mathfrak{v}(t)) \rangle
  = \mathbf{0}
\end{split}
\end{equation}
so that the first noise term in Eq.\ \eqref{eq:fokker-planck_derivation_langevin_inserted} vanishes. The second noise term reduces to its trace since each component of $\mathbf{N}$ is correlated only with itself, i.e., $\langle N_i (t) N_j (t) \rangle = \delta_{ij}$.

We thus find the \textit{inertial Fokker-Planck equation}
\begin{equation}
\begin{split}
    \frac{\partial P}{\partial t}
    &=\Big(
        - \mathbf{v}^\mathrm{T} \nabla_\mathbf{x}
        - \boldsymbol{\omega}^\mathrm{T} \nabla_{\mathrm{R}}
        - \nabla_\mathfrak{v}^\mathrm{T} \mathcal{M}^{-1} \big(
        -\eta \mathcal{R}^{-1} \mathcal{H}
        \mathcal{R} \mathfrak{v} + \mathcal{R}^{-1} \mathbf{K}_0
        + \mathbf{K}_\mathrm{ext} +  \mathbf{K}_\mathrm{D}
        \big) \\
        &\!\qquad + \frac{1}{2} \Tr\big((\sigma \mathcal{R} \mathcal{M}^{-1}) 
        \nabla_{\mathfrak{v}} \nabla_{\mathfrak{v}}^\mathrm{T} (\mathcal{M}^{-1}  
        \mathcal{R}^{-1} \sigma)\big)
     \Big) P \,,
\end{split}
\label{eq:fokker-planck}
\end{equation}
where $\Tr$ denotes the trace. This Fokker-Planck equation constitutes the second main result of this work.

In the special case of a spherical particle without self-propulsion in an external potential $\Phi(\mathbf{x})$, the inertial Fokker-Planck equation simplifies to the Kramers equation mentioned in Ref.\ \cite{Dhont1996},
\begin{equation}
\begin{split}
    \frac{\partial P}{\partial t}
    =\Big(
        - \mathbf{v}^\mathrm{T} \nabla_\mathbf{x}
        + \nabla_\mathbf{v}^\mathrm{T} m^{-1}
            \big(
                  \gamma \mathbf{v}
                + (\nabla_\mathbf{x} \Phi)
                - \mathbf{F}_\mathrm{D}
            \big)
        + m^{-2} \gamma k_\mathrm{B} T \nabla_\mathbf{v}^\mathrm{T} \nabla_\mathbf{v}
     \Big) P \,,
\end{split}
\label{eq:kramers}
\end{equation}
where the friction coefficient of the sphere $\gamma$ replaces the product of viscosity $\eta$ and hydrodynamic resistance matrix $\mathcal{H}$, i.e., $\eta \mathcal{H} \to \gamma$, and the rotational degrees of freedom are no longer relevant. 
\end{widetext}

We finish this section with a derivation of the drift term $\mathbf{K}_\mathrm{D}$ in the Langevin equation \eqref{eq:langevin} from the Fokker-Planck equation \eqref{eq:fokker-planck}.
In thermodynamic equilibrium, when the internal force is $\mathbf{K}_0 = \mathbf{0}$, the probability density is the time-independent Maxwell-Boltzmann distribution
\begin{equation}
	P_\mathrm{eq}(\mathfrak{x},\mathfrak{v}) =
	\frac{1}{Z}\exp{\!\bigg(-\beta \Big(\frac{1}{2}\mathfrak{v}^\mathrm{T}\mathcal{M}\mathfrak{v}
	- \int_{\mathfrak{x}_0}^\mathfrak{x} \mathbf{K}_\mathrm{ext} \,\mathrm{d}\mathfrak{x} \Big)}\bigg)\,,
	\label{eq:maxwell-boltzmann_distribution}
\end{equation}
where $Z$ denotes the partition function and $\mathfrak{x}_0$ the initial generalized position. The Fokker-Planck equation \eqref{eq:fokker-planck} lets us rewrite the time derivative of $P_\mathrm{eq}$ in terms of derivatives of generalized position $\mathfrak{x}$ and velocity $\mathfrak{v}$. 
We then obtain the condition
\begin{equation}
\begin{split}
	0 &= \frac{\partial P_\mathrm{eq}}{\partial t} \\
	  &= \Big(
        - \mathbf{v}^\mathrm{T} \nabla_\mathbf{x}
        - \boldsymbol{\omega}^\mathrm{T} \nabla_\mathrm{R} \\
        &\!\qquad - \nabla_\mathfrak{v}^\mathrm{T} \mathcal{M}^{-1} \big(
        -\eta \mathcal{R}^{-1} \mathcal{H}
        \mathcal{R} \mathfrak{v}
        + \mathbf{K}_\mathrm{ext} +  \mathbf{K}_\mathrm{D}
        \big) \\
        &\!\qquad + \frac{1}{2} \Tr\big((\sigma \mathcal{R} \mathcal{M}^{-1}) 
        \nabla_{\mathfrak{v}} \nabla_{\mathfrak{v}}^\mathrm{T} (\mathcal{M}^{-1}  
        \mathcal{R}^{-1} \sigma)\big)
     \Big) P_\mathrm{eq}\,,
\label{eq:drift_derivation_condition_prior}
\end{split}
\end{equation}
from which we aim to determine the drift term $\mathbf{K}_\mathrm{D}$. To do that, we will treat Eq.\ \eqref{eq:drift_derivation_condition_prior} term by term.

For the first one, we need to calculate
\begin{equation}
\begin{split}
	- \mathbf{v}^\mathrm{T} \nabla_\mathbf{x} P_\mathrm{eq}
	&= - \mathbf{v}^\mathrm{T} \nabla_\mathbf{x} \frac{1}{Z} \\
	&\quad\, \exp{\!\bigg(-\beta \Big(\frac{1}{2}\mathfrak{v}^\mathrm{T}\mathcal{M}\mathfrak{v}
	- \int_{\mathfrak{x}_0}^\mathfrak{x} \mathbf{K}_\mathrm{ext} \,\mathrm{d}\mathfrak{x} \Big)}\bigg)\,.
    \raisetag{10.2ex}
\end{split}
\end{equation}
The kinetic energy $\mathfrak{v}^\mathrm{T}\mathcal{M}\mathfrak{v} / 2$ does not directly depend on $\mathfrak{x}$, whereas the integral over the external generalized force $\mathbf{K}_\mathrm{ext}$ is in fact the sum of integrated force and torque,
\begin{equation}
	\int_{\mathfrak{x}_0}^\mathfrak{x} \mathbf{K}_\mathrm{ext} \, \mathrm{d}\mathfrak{x}
	= \int_{\mathbf{x}_0}^\mathbf{x} \mathbf{F}_\mathrm{ext} \, \mathrm{d}\mathbf{x}
	+ \int_{\mathrm{R}_0}^{\mathrm{R}} \mathbf{T}_\mathrm{ext} \, \mathrm{d}\mathrm{R}
\end{equation}
with $\mathfrak{x}_0 = (\mathbf{x}_0, \mathrm{R}_0)^\mathrm{T}$, so that 
\begin{equation}
\begin{split}
	- \mathbf{v}^\mathrm{T} \nabla_\mathbf{x} P_\mathrm{eq} 
	&= - \mathbf{v}^\mathrm{T} \beta P_\mathrm{eq} \nabla_\mathbf{x}
	\int_{\mathbf{x}_0}^\mathbf{x} \mathbf{F}_\mathrm{ext} \,\mathrm{d}\mathbf{x} \\
	&= - \beta \mathbf{v}^\mathrm{T} \mathbf{F}_\mathrm{ext} P_\mathrm{eq} \,.
\end{split}
\label{eq:drift_derivation_force}
\end{equation}
Similarly, the rotational gradient operator $\nabla_{\mathrm{R}}$ in the second term of Eq.\ \eqref{eq:drift_derivation_condition_prior} acts only on the integrated torque in the exponent of $P_\mathrm{eq}$:
\begin{equation}
\begin{split}
	- \boldsymbol{\omega}^\mathrm{T} \nabla_{\mathrm{R}} P_\mathrm{eq}
	&= - \boldsymbol{\omega}^\mathrm{T} \beta  P_\mathrm{eq} \nabla_{\mathrm{R}}
	\int_{\mathrm{R}_0}^{\mathrm{R}} \mathbf{T}_\mathrm{ext} \,\mathrm{d}\mathrm{R} \\
	&= - \beta \boldsymbol{\omega}^\mathrm{T} \mathbf{T}_\mathrm{ext} P_\mathrm{eq} \,.
\end{split}
\label{eq:drift_derivation_torque}
\end{equation}

The third term, which stems from the Langevin equation, requires some attention to detail.
By the product rule for a scalar function $f$ and a matrix $\mathcal{A}$,
\begin{equation}
	\nabla_\mathfrak{v}^\mathrm{T} \mathcal{A} \mathfrak{v} f(\mathfrak{v})
	= f(\mathfrak{v}) \nabla_\mathfrak{v}^\mathrm{T} \mathcal{A} \mathfrak{v}
	+ (\nabla_\mathfrak{v} f(\mathfrak{v}))^\mathrm{T} \mathcal{A} \mathfrak{v}\,.
\end{equation}
Explicit calculation shows that if $\mathcal{A}$ is symmetric,
\begin{equation}
	\nabla_\mathfrak{v}^\mathrm{T} \mathcal{A} \mathfrak{v} = \Tr(\mathcal{A})\,.
\end{equation}
In particular, with $f(\mathfrak{v}) \equiv P_\mathrm{eq}$ we find
\begin{equation}
	\nabla_\mathfrak{v}^\mathrm{T} \mathcal{A} \mathfrak{v} P_\mathrm{eq}
	= P_\mathrm{eq} \Tr(\mathcal{A})
	- \beta \mathfrak{v}^\mathrm{T} \mathcal{M} P_\mathrm{eq} \mathcal{A} \mathfrak{v}
	\label{eq:drift_derivation_product_rule_lemma}
\end{equation}
since $\mathcal{M}$ is symmetric.
We assume that $\mathbf{K}_\mathrm{ext}$ is partially independent of $\mathfrak{v}$ and use Eq.\ \eqref{eq:drift_derivation_product_rule_lemma} with $\mathcal{A} \equiv - \eta \mathcal{M}^{-1} \mathcal{R}^{-1} \mathcal{H} \mathcal{R}$. It follows that
\begin{equation}
\begin{split}
	- &\nabla_\mathfrak{v}^\mathrm{T} \mathcal{M}^{-1} \big(
        -\eta \mathcal{R}^{-1} \mathcal{H}
        \mathcal{R} \mathfrak{v}
        + \mathbf{K}_\mathrm{ext} +  \mathbf{K}_\mathrm{D}
        \big) P_\mathrm{eq} \\
    =& - P_\mathrm{eq} \big(-\eta  \Tr( \mathcal{M}^{-1}
        \mathcal{R}^{-1} \mathcal{H}
        \mathcal{R}) + \nabla_\mathfrak{v}^\mathrm{T} \mathcal{M}^{-1} \mathbf{K}_\mathrm{D} \big) \\
       & - (- \beta) \mathfrak{v}^\mathrm{T} \mathcal{M} P_\mathrm{eq} \big(\mathcal{M}^{-1} (
        -\eta \mathcal{R}^{-1} \mathcal{H}
        \mathcal{R} \mathfrak{v}
        + \mathbf{K}_\mathrm{ext} +  \mathbf{K}_\mathrm{D}
        )\big) \\
    =& \Big(- \big(-\eta  \Tr( \mathcal{M}^{-1}
        \mathcal{R}^{-1} \mathcal{H}
        \mathcal{R}) + \nabla_\mathfrak{v}^\mathrm{T} \mathcal{M}^{-1} \mathbf{K}_\mathrm{D} \big) \\
       & + \beta \mathfrak{v}^\mathrm{T} \big(
        -\eta \mathcal{R}^{-1} \mathcal{H}
        \mathcal{R} \mathfrak{v}
        + \mathbf{K}_\mathrm{ext} +  \mathbf{K}_\mathrm{D}
        \big) \Big) P_\mathrm{eq}\,.
\raisetag{3.8ex}
\label{eq:drift_derivation_langevin_term_result}
\end{split}
\end{equation}
We observe that the external generalized force $\mathbf{K}_\mathrm{ext}$ in the last line cancels out the contributions from the external force \eqref{eq:drift_derivation_force} and torque \eqref{eq:drift_derivation_torque} since

\begin{equation}
	\beta \mathfrak{v}^\mathrm{T} \mathbf{K}_\mathrm{ext}  P_\mathrm{eq}
	= \beta (\mathbf{x}^\mathrm{T} \mathbf{K}_\mathrm{ext}
	+ \boldsymbol{\omega}^\mathrm{T} \mathbf{T}_\mathrm{ext}) P_\mathrm{eq}\,.
\end{equation}

The last term,
\begin{equation}
	 \frac{1}{2} \Tr\big((\sigma \mathcal{R} \mathcal{M}^{-1}) 
        \nabla_{\mathfrak{v}} \nabla_{\mathfrak{v}}^\mathrm{T} (\mathcal{M}^{-1}  
        \mathcal{R}^{-1} \sigma)\big) P_\mathrm{eq}\,,
\end{equation}
arises through the stochastic force and torque. The trace is cyclic, which means that the matrices inside can be reordered to
\begin{equation}
	 \frac{1}{2} \Tr\big(
	 (\mathcal{M}^{-1} \mathcal{R}^{-1} \sigma)
     (\sigma \mathcal{R} \mathcal{M}^{-1}) 
     \nabla_{\mathfrak{v}} \nabla_{\mathfrak{v}}^\mathrm{T} P_\mathrm{eq} \big)\,.
\label{eq:drift_derivation_stochastic_term}
\end{equation}
This allows us to apply the operators $\nabla_{\mathfrak{v}}$ on $P_\mathrm{eq}$, which, being scalar, can be moved inside the trace. The Hessian matrix of $P_\mathrm{eq}$ yields, after some regrouping,
\begin{equation}
	\nabla_{\mathfrak{v}} \nabla_{\mathfrak{v}}^\mathrm{T} P_\mathrm{eq}
	= - \beta \mathcal{M} P_\mathrm{eq} + \beta ^2 \mathcal{M} \mathfrak{v} \mathfrak{v}^\mathrm{T} \mathcal{M} P_\mathrm{eq} \,.
\label{eq:drift_derivatin_hessian}
\end{equation}
Note that both expressions on the right-hand side are still matrices, as $\mathfrak{v} \mathfrak{v}^\mathrm{T}$ is an outer product. We insert Eq.\ \eqref{eq:drift_derivatin_hessian} into Eq.\ \eqref{eq:drift_derivation_stochastic_term} and use again the cyclicity and linearity of the trace, which gives
\begin{equation}
\begin{split}
	&\frac{1}{2} \Tr\Big(
	 (\mathcal{M}^{-1} \mathcal{R}^{-1} \sigma)
     (\sigma \mathcal{R} \mathcal{M}^{-1}) \\
     &\qquad\;\; \big(- \beta \mathcal{M} + \beta ^2 (\mathcal{M} \mathfrak{v}) (\mathfrak{v}^\mathrm{T} \mathcal{M}) \big) \Big) P_\mathrm{eq} \\
     &= \beta \frac{1}{2} \Tr\Big(
	 (\mathcal{M}^{-1} \mathcal{R}^{-1} \sigma)
     (\sigma \mathcal{R}) 
     (- \mathrm{I} + \beta \mathfrak{v} \mathfrak{v}^\mathrm{T} \mathcal{M}) \Big) P_\mathrm{eq} \\
     &= \beta \frac{1}{2} \Big( - \Tr\big(
	 \mathcal{M}^{-1} \mathcal{R}^{-1} \sigma^2 \mathcal{R} \big) 
     + \beta \Tr\big(\mathcal{R}^{-1} \sigma^2 \mathcal{R} \mathfrak{v} \mathfrak{v}^\mathrm{T} \big) \Big) P_\mathrm{eq} \,.
\raisetag{12.5ex}
\end{split}
\end{equation}
Using that the variation is $\sigma^2 = \mathcal{H} \frac{2\eta}{\beta}$, we obtain
\begin{equation}
\begin{split}
    \dots = \eta \Big( - \Tr\big(
	 \mathcal{M}^{-1} \mathcal{R}^{-1} \mathcal{H} \mathcal{R} \big) 
     + \beta \Tr\big(\mathcal{R}^{-1}  \mathcal{H} \mathcal{R} \mathfrak{v} \mathfrak{v}^\mathrm{T} \big) \Big) P_\mathrm{eq} \,.
\end{split}
\label{eq:drift_derivation_stochastic_term_result}
\end{equation}

Finally, we may combine the results \eqref{eq:drift_derivation_langevin_term_result} and \eqref{eq:drift_derivation_stochastic_term_result} with the requirement \eqref{eq:drift_derivation_condition_prior} and simplify:
\begin{equation}
\begin{split}
	0 = & \Big(\eta \Tr\big( \mathcal{M}^{-1} \mathcal{R}^{-1} \mathcal{H} \mathcal{R}\big)
        - \nabla_\mathfrak{v}^\mathrm{T} \mathcal{M}^{-1} \mathbf{K}_\mathrm{D} \\
        &\;\; - \beta \eta \mathfrak{v}^\mathrm{T} \mathcal{R}^{-1} \mathcal{H} 
            \mathcal{R} \mathfrak{v} + \beta \mathfrak{v}^\mathrm{T} \mathbf{K}_\mathrm{D} \Big) P_\mathrm{eq} \\
        &+ \Big( - \eta \Tr\big(
            \mathcal{M}^{-1} \mathcal{R}^{-1} \mathcal{H} \mathcal{R} \big) \\
        &\quad\;\; + \beta \eta \Tr\big(\mathcal{R}^{-1}  \mathcal{H} \mathcal{R} \mathfrak{v} \mathfrak{v}^\mathrm{T} \big) \Big) P_\mathrm{eq}\,.
\end{split}
\end{equation}
The traces in the first and second to last term on the right-hand side cancel out, and $P_\mathrm{eq} = 0$ is no relevant solution, which leaves
\begin{equation}
\begin{split}
	0 = &- \nabla_\mathfrak{v}^\mathrm{T} \mathcal{M}^{-1} \mathbf{K}_\mathrm{D} + \beta \mathfrak{v}^\mathrm{T} \mathbf{K}_\mathrm{D} \\
        &-\beta \eta \mathfrak{v}^\mathrm{T} \mathcal{R}^{-1} \mathcal{H}
            \mathcal{R} \mathfrak{v} + \beta \eta \Tr\big(\mathcal{R}^{-1}  \mathcal{H} \mathcal{R} \mathfrak{v} \mathfrak{v}^\mathrm{T} \big) \,.
\label{eq:drift_derivation_condition_last_step}
\end{split}
\end{equation}
The last step is to use the property that the trace of the outer product of real column matrices $\mathbf{a}$ and $\mathbf{b}$ is equal to their scalar product, i.e.,
\begin{equation}
	 \Tr(\mathbf{b} \mathbf{a}^\mathrm{T}) = \mathbf{a}^\mathrm{T} \mathbf{b}\,.
\label{eq:drift_derivation_trace_column_matrices}
\end{equation}
From this we see that
\begin{equation}
\begin{split}
	\beta \eta \Tr\big( (\mathcal{R}^{-1} \mathcal{H} \mathcal{R} \mathfrak{v}) \mathfrak{v}^\mathrm{T} \big)
	&= \beta \eta \mathfrak{v}^\mathrm{T} (\mathcal{R}^{-1} \mathcal{H} \mathcal{R} \mathfrak{v})\,.
\end{split}
\end{equation}
Condition \eqref{eq:drift_derivation_condition_last_step} thus becomes
\begin{equation}
	\nabla_\mathfrak{v}^\mathrm{T} \mathcal{M}^{-1} \mathbf{K}_\mathrm{D}
	= \beta \mathfrak{v}^\mathrm{T} \mathbf{K}_\mathrm{D}
\end{equation}
to which $\mathbf{K}_\mathrm{D} = \mathbf{0}$ is a possible solution\footnote{The drift term $\mathbf{K}_\mathrm{D} = \mathbf{c} \exp{(+(\beta / 2)\, \mathfrak{v}^\mathrm{T}\mathcal{M}\mathfrak{v})}$ solves the equation as well for $\mathbf{c} \in \mathbb{R}^3\backslash\{\mathbf{0}\}$, yet leads to quickly diverging solutions to the Langevin equation and does not converge to the overdamped case when inertia is neglected ($\mathcal{M}\to 0$) since $\exp{(+(\beta / 2)\, \mathfrak{v}^\mathrm{T}\mathcal{M}\mathfrak{v})} \rightarrow 1$ for vanishing $\mathcal{M}$, while $\mathcal{M}^{-1}$ tends to infinity on the right-hand side of the Langevin equation \eqref{eq:langevin}. This second solution is therefore implausible.}.
This stands in contrast to the overdamped case, for which the drift term is $\mathbf{K}_\mathrm{D,B} = (\beta \mathcal{D}(\mathfrak{x}))^{-1} (\nabla_\mathfrak{x}^\mathrm{T} \mathcal{D}(\mathfrak{x}))^\mathrm{T}$.

The observation that no drift term is necessary for inertial Brownian motion agrees with all stochastic integral conventions coinciding, such as the It\^o, Stratonovich, and isothermal (also called H\"anggi-Klimontovich) integral convention \cite{Kampen1981}. This freedom of choice roots in that the conventions only differ in their point of evaluation within a small time step $\Delta t$, over which the noise is integrated, and thus converge to the same integral for $\Delta t \to 0$, albeit with different speeds \cite{LauL2007, FaragoGL2014}. The isothermal convention happens to imply a vanishing drift term. One can reason that the drift term is consequently zero for all conventions in the inertial case since all conventions should yield the same result with inertia.

\section{\label{sec:numerical_methods}Numerical methods}
This section shows how to simulate the Brownian motion of a self-propelled asymmetric particle numerically. We begin with a nondimensionalization, go on to present the discrete inertial and overdamped Langevin equations, and conclude with some technical remarks.

\subsection{\label{sec:nondimensionalization}Nondimensionalization}
To nondimensionalize the Langevin equation \eqref{eq:langevin}, we introduce the characteristic scales $l_\mathrm{c}$ for length, $m_\mathrm{c}$ for mass, and $t_\mathrm{c}$ for time. Every variable can then be transformed into a dimensionless counterpart, denoted by a hat on top, e.g.,
$t = t_\mathrm{c} \hat{t}$.
In the following, we list the nondimensionalization of relevant quantities. The particle state is given by 
\begin{align}
    \mathbf{x} &= l_\mathrm{c} \mathbf{\hat x}\,, \quad
    \mathcal{R} = \mathcal{\hat R}\,, \quad
    \mathfrak{v} = \frac{l_\mathrm{c}}{t_\mathrm{c}} \,\mathrm{diag}(1, 1 / l_\mathrm{c})\, \mathfrak{\hat v} \,,
\label{eq:nondimen_scales}
\end{align}
where $\diag(a, b)$ is defined here as
\begin{equation}
    \mathrm{diag}(a, b) = \begin{pmatrix} a \mathrm{I} & 0 \\ 0 & b \mathrm{I}  \end{pmatrix}.
\end{equation}
The generalized mass is
\begin{equation}
    \mathcal{M} = \mathrm{diag}(m_\mathrm{c}, m_\mathrm{c} l_\mathrm{c}^2) \mathcal{\hat M}\,,
\label{eq:nondimen_mass}
\end{equation}
the hydrodynamic resistance matrix is
\begin{equation}
    \mathcal{H} = \begin{pmatrix} l_\mathrm{c} \mathrm{ \hat K} & l_\mathrm{c}^2 \mathrm{\hat C}_\mathrm{S}^\mathrm{T}    \\
    l_\mathrm{c}^2 \mathrm{\hat C}_\mathrm{S} & l_\mathrm{c}^3 \hat{\Omega}_\mathrm{S} \end{pmatrix}
  = l_\mathrm{c}
    \,\mathrm{diag}(1, l_\mathrm{c})
    \mathcal{\hat H}
    \,\mathrm{diag}(1, l_\mathrm{c})\,,
\label{eq:nondimen_hydrodynamic}
\end{equation}
and the diffusion tensor is consequently
\begin{equation}
    \mathcal{D}(\mathcal{R}) =
    \frac{l_\mathrm{c}^2}{t_\mathrm{c}} \,\mathrm{diag}(1, 1/l_\mathrm{c}) \mathcal{\hat D}(\mathcal{R}) \,\mathrm{diag}(1, 1/l_\mathrm{c})\,.
\label{eq:}
\end{equation}
Finally, a generalized force is nondimensionalized as
\begin{equation}
      \mathbf{K}_\mathrm{\dotso}(\mathbf{x},\mathcal{R})
      = \frac{m_\mathrm{c} l_\mathrm{c}}{t_\mathrm{c}^2} \,\mathrm{diag}(1, l_\mathrm{c})\, \mathbf{\hat K}_\mathrm{\dotso}(\mathbf{\hat x},\mathcal{R})\,.
\label{eq:nondimen_genforce}
\end{equation}

Substituting Eqs.\ \eqref{eq:nondimen_scales} and \eqref{eq:nondimen_mass}-\eqref{eq:nondimen_genforce} into the Langevin equation \eqref{eq:langevin}, we get
\begin{equation}
\begin{split}
    &m_\mathrm{c} \,\mathrm{diag}(1, l_\mathrm{c}^2) \mathcal{\hat M} \frac{\partial}{\partial (t_\mathrm{c} \hat t)}
        \frac{l_\mathrm{c}}{t_\mathrm{c}} \,\mathrm{diag}(1, 1 / l_\mathrm{c})\,   \mathfrak{\hat v} \\
    &= - \eta \mathcal{R}^{-1} l_\mathrm{c}
        \,\mathrm{diag}(1, l_\mathrm{c}) \mathcal{\hat H} \,\mathrm{diag}(1, l_\mathrm{c}) \mathcal{R} \frac{l_\mathrm{c}}{t_\mathrm{c}} \,\mathrm{diag}(1, 1 / l_\mathrm{c})\,   \mathfrak{\hat v} \\
    &\quad\, + \frac{m_\mathrm{c} l_\mathrm{c}}{t_\mathrm{c}^2} \,\mathrm{diag}(1, l_\mathrm{c})\, \mathbf{\hat K}_\mathrm{rest}(\mathbf{\hat x},\mathcal{R})\,,
\raisetag{4.55ex}
\end{split}
\end{equation}
where $\mathbf{\hat K}_\mathrm{rest}$ comprises the remaining forces and torques, which have the same dimension, in
\begin{equation}
	\mathbf{K}_\mathrm{rest} = \mathcal{R}^{-1} \mathbf{K}_0 + \mathbf{K}_\mathrm{ext}
	+  \mathbf{K}_\mathrm{D} + \mathcal{R}^{-1} \mathbf{k}\,.
\end{equation}
Since the diagonal matrices $\diag(\dotsc,\dots)$ commute with the block diagonal matrices $\mathcal{M}$ and $\mathcal{R}$, the nondimensionalized Langevin equation simplifies to
\begin{equation}
  \mathcal{\hat M} \frac{\partial}{\partial \hat t} \mathfrak{\hat v}
  = -  \frac{l_\mathrm{c} t_\mathrm{c}}{m_\mathrm{c}} \eta \mathcal{R}^{-1} \mathcal{\hat H} \mathcal{R} \mathfrak{\hat v}
  + \mathbf{\hat K}_\mathrm{rest}(\mathbf{\hat x},\mathcal{R})\,.
\label{eq:nondimen_langevin}
\end{equation}

In a similar manner, substitution of Eqs.\ \eqref{eq:nondimen_scales} and \eqref{eq:nondimen_mass}-\eqref{eq:nondimen_genforce} into relation \eqref{eq:diffusion_tensor_definition} connecting diffusion tensor and hydrodynamic resistance matrix gives
\begin{equation}
\begin{split}
    & \eta \mathcal{R}^{-1} l_\mathrm{c}
        \,\mathrm{diag}(1, l_\mathrm{c}) \mathcal{\hat H} \,\mathrm{diag}(1, l_\mathrm{c}) \mathcal{R} \\
    &\!= \frac{1}{\beta} \frac{t_\mathrm{c}}{l_\mathrm{c}^2}
        \,\mathrm{diag}(1, l_\mathrm{c})\mathcal{\hat D}^{-1}(\mathcal{R}) \,\mathrm{diag}(1, l_\mathrm{c})\,,
\end{split}
\end{equation}
which reduces to
\begin{equation}
	\mathcal{R}^{-1}
  \mathcal{\hat H}\mathcal{R}
  = \frac{1}{\eta \beta} \frac{t_\mathrm{c}}{l_\mathrm{c}^3}
  \mathcal{\hat D}^{-1}(\mathcal{R})\,.
\label{eq:nondimen_diffusion}
\end{equation}

Finally, Eq.\ \eqref{eq:noise_variance}, which describes the auto-correlation of the noise, becomes
\begin{equation}
\begin{split}
    &\frac{m_\mathrm{c}^2 l_\mathrm{c}^2}{t_\mathrm{c}^4} \,\mathrm{diag}(1, l_\mathrm{c})
        \langle \hat{\mathbf{k}}(\hat t_1) \otimes \hat{\mathbf{k}}(\hat t_2) \rangle
        \,\mathrm{diag}(1, l_\mathrm{c}) \\
    &= l_\mathrm{c}
        \,\mathrm{diag}(1, l_\mathrm{c})
        \mathcal{\hat H}
        \,\mathrm{diag}(1, l_\mathrm{c})
        \frac{2 \eta}{\beta} \delta(t_\mathrm{c}\hat t_1 - \hat t_\mathrm{c} t_2)
\end{split}
\end{equation}
and simplifies to
\begin{equation}
	\langle \hat{\mathbf{k}}(\hat t_1) \otimes \hat{\mathbf{k}}(\hat t_2) \rangle
 	= \mathcal{\hat H}
    \frac{2 \eta}{\beta} \frac{t_\mathrm{c}^3}{m_\mathrm{c}^2 l_\mathrm{c}} \delta(\hat t_1 - \hat t_2)\,.
\label{eq:nondimen_correlation}
\end{equation}

The Eqs.\ \eqref{eq:nondimen_langevin}, \eqref{eq:nondimen_diffusion}, and \eqref{eq:nondimen_correlation} lead to a natural set of dimensionless parameters,
\begin{align}
	\mu &= \frac{l_\mathrm{c} t_\mathrm{c}}{m_\mathrm{c}} \eta \,, \label{eq:nondimen_mu} \\
	\nu &= \frac{1}{\eta \beta} \frac{t_\mathrm{c}}{l_\mathrm{c}^3} \,, \label{eq:nondimen_nu} \\
	\kappa &= \frac{\eta}{\beta} \frac{t_\mathrm{c}^3}{m_\mathrm{c}^2 l_\mathrm{c}}
	    = \mu^2 \nu \label{eq:nondimen_kappa} \,.
\end{align}

So far, the characteristic scales have been left unspecified. The size of the particle is a sensible choice for the characteristic length.
Since the particle's shape need not be spherical, it is reasonable to define $l_\mathrm{c}$ as the diameter of a sphere with the particle's volume $V$:
\begin{equation}
	l_\mathrm{c} = \sqrt[3]{\frac{6 V}{\pi}}\,.
\end{equation}
The obvious choice for the characteristic mass $m_\mathrm{c}$ is the particle's mass:
\begin{equation}
	m_\mathrm{c} = m\,.
\end{equation}
One degree of freedom remains in the choice of the characteristic scales, so we conveniently set $\mu = 1$, which defines the characteristic time scale as
\begin{equation}
t_\mathrm{c} = \frac{m_\mathrm{c}}{\eta l_\mathrm{c}}
\label{eq:nondimen_timescale}
\end{equation}
and leads to $\kappa = \nu$. This leaves $\nu$ as sole dimensionless scaling parameter, which
governs the relative strength of the generalized stochastic force $\mathbf{k}$.
A brief argument, given in \cref{sec:app:nondimen_interpretation_time_scale}, shows that $t_\mathrm{c}$ is the time after which a free spherical particle of radius $l_\mathrm{c} / (6 \pi)$ with initial velocity $v_0$ has slowed down to $v_0 / \mathrm{e}$.

With the definitions \eqref{eq:nondimen_mu}-\eqref{eq:nondimen_kappa}, we summarize the nondimensionalized equations as
\begin{align}
    \begin{split}
    \mathcal{\hat M} \frac{\partial}{\partial \hat t} \mathfrak{\hat v}
    &= - \mathcal{R}^{-1} \mathcal{\hat H} \mathcal{R} \mathfrak{\hat v}
    + \mathcal{R}^{-1} \mathbf{\hat K}_0 + \mathbf{\hat K}_\mathrm{ext}
    +  \mathbf{\hat K}_\mathrm{D} \\
    &\quad\,+ \mathcal{R}^{-1} \mathbf{\hat k} \,, 
    \raisetag{3.3ex}
    \label{eq:nondimen_langevin_mu}
    \end{split} \\
    \mathcal{R}^{-1} \mathcal{\hat H} \mathcal{R}
    &= \nu \mathcal{\hat D}^{-1} \,,
    \label{eq:nondimen_diffusion_nu} \\
    \langle \hat{\mathbf{k}}(\hat t_1) \otimes \hat{\mathbf{k}}(\hat t_2) \rangle
    &= 2 \nu \mathcal{\hat H} \delta(\hat t_1 - \hat t_2) \,.
    \label{eq:nondimen_correlation_kappa}
\end{align}
The hat for nondimensionalized variables will be omitted in the following.

\subsection{\label{sec:discrete_langevin_equation}Discretization}
To solve the Langevin equation \eqref{eq:nondimen_langevin_mu} numerically, it needs to be discretized. This is done via the Euler-Maruyama method, which is an extension of the Euler method to stochastic differential equations. The Euler-Maruyama method is an explicit first-order scheme \cite{KloedenP2006}.

\subsubsection{Discrete inertial Langevin equation}
We discretize time as $t_n=n\Delta t$ with time step size $\Delta t$ and denote a quantity $X$ at time step $n$ as $X_n$.
The nondimensionalized discrete Langevin equation that approximates the $(n+1)$th step of the generalized velocity $\mathfrak{v}_{n+1}$ based on the particle state in the $n$th step is then given by
\begin{equation}
\begin{split}
    \mathfrak{v}_{n+1} &= \mathfrak{v}_n \\
    &\quad + \mathcal{M}^{-1}
        \big(
            - \mathcal{R}_n^{-1} \mathcal{H} \mathcal{R}_n \mathfrak{v}_n
            + \mathcal{R}_n^{-1} \mathbf{K}_{0,n}
            + \mathbf{K}_\mathrm{ext,n}
        \big) \Delta t \\
    &\quad + \mathcal{M}^{-1} \mathcal{R}_n^{-1} \sigma \mathbf{N}_n \sqrt{\Delta t} \,,
\label{eq:discrete_langevin}
\raisetag{3.5ex}
\end{split}
\end{equation}
where $\mathbf{N}_n$ is a vector of statistically independent\footnote{The statistical independence applies both to different elements of the vector $\mathbf{N}_n$ and to the elements at different time steps $n$.} variables with standard normal distribution $\mathcal{N}(0,1)$.

The position $\mathbf{x}_{n+1}$ is obtained through
\begin{equation}
	\mathbf{x}_{n+1} = \mathbf{x}_{n} + \mathbf{v}_{n} \Delta t 
\end{equation}
and the orientation matrix $\mathrm{R}_{n+1}$ through
\begin{equation}
	\mathrm{R}_{n+1} = \mathrm{R}_{n} (\mathrm{I} + \mathrm{W}_{n} \Delta t)\,,
\label{eq:discrete_rotation}
\end{equation}
where $\mathrm{W}$ is a skew-symmetric tensor, namely the Hodge dual $\star$ of the angular velocity, defined by
\begin{equation}
	\mathrm{W} = \star \boldsymbol{\omega}
	= \begin{pmatrix*}[r]
    0        & \omega_3  & -\omega_2 \\
   -\omega_3 &  0        &  \omega_1 \\ 
    \omega_2 & -\omega_1 &  0
  \end{pmatrix*}.
\end{equation}
To ensure numerical stability, the rotation matrix $\mathrm{R}_n$ is orthonormalized with the Gram-Schmidt process in each step.
Equations \eqref{eq:discrete_langevin}-\eqref{eq:discrete_rotation} constitute the numerical scheme of inertial Brownian motion. See \cref{sec:app:motivation_of_rotation} for a motivation of Eq.\ \eqref{eq:discrete_rotation}.

\subsubsection{Discrete overdamped Langevin equation}
The overdamped Brownian motion described by the Langevin equation \eqref{langevin_overdamped} differs in its nondimensionalized discretization from the inertial case only in that the generalized velocity is given by
\begin{equation}
\begin{split}
	\mathfrak{v}_{n+1} &= \mathcal{D}_n
	\big(\mathcal{R}_n^{-1} \mathbf{K}_{0,n} + \mathbf{K}_\mathrm{ext,n}
	+ \mathcal{R}_n^{-1}  \sigma \mathbf{N}_n / \sqrt{\Delta t} \big) \\
	&\quad + \big(\mathbf{\nabla}_{\mathbf{\mathfrak{x}}}^\mathrm{T} \mathcal{D}\big)^\mathrm{T}_n\,.
\label{eq:discrete_langevin_overdamped}
\end{split}
\end{equation}

One difficulty remains, however, with the overdamped scheme: How to implement the derivative of the diffusion tensor $\mathbf{\nabla}_{\mathbf{\mathfrak{x}}} \mathcal{D}(\mathbf{\mathfrak{x}})$ numerically?
Our starting point is to use the nondimensionalized relation \eqref{eq:nondimen_diffusion_nu} that links diffusion tensor and hydrodynamic resistance matrix to get
\begin{equation}
	\nabla_\mathfrak{x} ^\mathrm{T}\mathcal{D} = \nu
	\big(
		\nabla_\mathbf{x}^\mathrm{T} , \nabla_\mathrm{R}^\mathrm{T}
	\big)
	\begin{pmatrix}
		\mathrm{R}^{-1} (\mathrm{H}^{-1})_{11} \mathrm{R}
		& \mathrm{R}^{-1} (\mathrm{H}^{-1})_{12} \mathrm{R} \\
		\mathrm{R}^{-1} (\mathrm{H}^{-1})_{21} \mathrm{R}
		& \mathrm{R}^{-1} (\mathrm{H}^{-1})_{22} \mathrm{R}
	\end{pmatrix} ,
\end{equation}
where $(\mathrm{H}^{-1})_{ij}$ means the $3\times3$--dimensional block matrix in the $i$th row and $j$th column of the  $6\times6$--dimensional matrix $\mathcal{H}^{-1}$ for $i,j = 1, 2$.
The gradient $\nabla_\mathbf{x}$ vanishes since the rotation matrix $\mathrm{R}$ depends solely on the Euler angles $\boldsymbol{\varpi}$. Therefore, we have
\begin{equation}
	\nabla_\mathfrak{x}^\mathrm{T} \mathcal{D} = \nu
	\big(
		\nabla_\mathrm{R}^\mathrm{T}
		\mathrm{R}^\mathrm{T} (\mathrm{H}^{-1})_{21} \mathrm{R} ,
		\nabla_\mathrm{R}^\mathrm{T}
		\mathrm{R}^\mathrm{T} (\mathrm{H}^{-1})_{22} \mathrm{R}
	\big).
\end{equation}
Explicit calculation with the definition \eqref{eq:rotation_matrix_definition} of the rotation matrix shows that for any $3\times3$--matrix $\mathrm{A}$ (see \cref{sec:app:overdamped_drift_term_calculation})
\begin{equation}
\begin{split}
	\nabla_\mathrm{R}^\mathrm{T} \mathrm{R}^\mathrm{T} \mathrm{A} \mathrm{R} 
	&= (
	\mathrm{A}_{23} - \mathrm{A}_{32} , - (\mathrm{A}_{13} - \mathrm{A}_{31}) , \mathrm{A}_{12} - \mathrm{A}_{21}
	) \, \mathrm{R} \\
	&= (\mathrm{R}^\mathrm{T} \Delta \mathrm{A})^\mathrm{T}\,,
\raisetag{3.25ex}
\end{split}
\end{equation}
where
\smash{$\Delta \mathrm{A} =
(
	\mathrm{A}_{23} - \mathrm{A}_{32}  ,
 - (\mathrm{A}_{13} - \mathrm{A}_{31}) ,
    \mathrm{A}_{12} - \mathrm{A}_{21}
)^\mathrm{T}$}
is defined as the difference of the off-diagonal elements \mbox{of $\mathrm{A}$}. It is then obvious that $\nabla_\mathrm{R} \mathrm{R}^\mathrm{T} (\mathrm{H}^{-1})_{22} \mathrm{R} = \mathbf{0}$ since the inverse of a symmetric matrix such as $\mathcal{H}$ is itself symmetric and, consequently, $(\mathrm{H}^{-1})_{22}$ is symmetric, too.
We have thus found a concise formula that calculates the contribution to velocity due to the overdamped drift term:
\begin{equation}
	\big(\nabla_\mathfrak{x}^\mathrm{T} \mathcal{D}\big)^\mathrm{T} = \nu \begin{pmatrix}
		\mathrm{R}^\mathrm{T} \Delta ((\mathrm{H}^{-1})_{21}) \\ \mathbf{0}
	\end{pmatrix}.
\end{equation}

\subsection{\label{sec:simulation_parameters}Simulation parameters}
We implemented the Euler-Maruyama method in \textsc{Fortran} with a wrapper in \textsc{Python} \cite{SI}.

The hydrodynamic resistance matrix $\mathcal{H}$ needs to be randomly generated to classify the trajectories for arbitrarily shaped particles. Not every matrix, however, is physically sensible. A method to generate physical matrices is to begin with an orthotropic particle, which means that $\mathcal{H}$ is diagonal. We then translate and rotate the reference frame as described in Ref.\ \cite{VossW2018}. Similarly, we generate a mass $m$ and the principal moments for the moment of inertia tensor, which is then rotated randomly, to obtain a random generalized mass $\mathcal{M}$.

We specify here the parameters for three simulations presented in sections \ref{sec:hydrodynamic_friction_tensor} and \ref{sec:simulation_and_analytical_solution}. We choose $m_\mathrm{c} = \SI{1}{\pico\gram}$, $l_\mathrm{c} = \SI{1}{\micro\meter}$, and $t_\mathrm{c} = \SI{1}{\micro\second}$ as characteristic scales. All following values in this section are stated in these base units. The viscosity in all simulations is $\eta = \SI{1}{\milli\pascal\second}$, which corresponds to water at room temperature. The initial conditions are $\mathbf{x}_0 = (0, 0, 0)^\mathrm{T}$, $\boldsymbol{\varpi}_0 = (0, 0, 0)^\mathrm{T}$, and $\mathfrak{v}_0 = \mathbf{0}$.

In the first simulation, the particle is an irregular trimer. \cref{sec:app:numerical_hydrodynamic_friction_tensor} lists its $\mathcal{H}$ and $\mathcal{M}$. The self-propulsion is defined by $\mathbf{K}_0 = (0,0,3,1,1,1)^\mathrm{T} \cdot 10^{-3}$.
Noisy trajectories at $T = \SI{300}{\kelvin}$ were simulated for $\SI{1}{\micro\second}$ for log-equally distributed numbers of steps $N$ from $32$ up to $10^5$.

The second simulation tests whether all particles at $T=\SI{0}{\kelvin}$ move along circular helices if there are no external force and torque. To check this proposition, we generate $1000$ particles with random hydrodynamic resistance matrix $\mathcal{H}$, generalized mass $\mathcal{M}$, and internal force and torque $\mathbf{K}_0$. We must, of course, restrict ourselves to a finite interval from which to draw the values for these parameters. We set the intervals to $[0.1, 10]$ for the principal values of $\mathcal{H}$, to $[0.1, 1000]$ for the principal values of $\mathcal{M}$, and to $[0.1, 10] \cdot 10^{-3}$ for the internal force and torque. These boundaries correspond roughly to typical sizes for microorganisms: \emph{Escherichia coli}, for example, is a rod-shaped bacterium that is about $\SI{2}{\micro\meter}$ long and $\SI{0,5}{\micro\meter}$ in diameter and has a mass of $\SI{1}{\pico\gram}$ \cite{DonarchieBV1976}. \mbox{\emph{E.\ coli}} has an average thrust force of $\SI{0.57}{\pico\newton}$ and a thrust torque of $0.5\,\si{\pico\newton\micro\meter}$ \cite{ChattopadhyayMYW2006}. The intervals cover the range from overdamped to highly inertial particles and allow for a variety of extreme shapes and methods of propulsion. If we drew the random values from a uniform distribution over the intervals above, the values would naturally fall mostly into the highest order of magnitude. Since we want to include extreme cases that are caused if the principal values differ by several orders of magnitude, we draw the values from a log-uniform distribution instead.

For each particle, the trajectory is simulated for $\SI{20}{\milli\second}$ with $N = 10^6$ steps. If the radius of translational velocity or angular velocity is not constant after some time, the simulation repeats with longer simulation time and smaller step size. These iterations are useful to save computation time because some particles have stable trajectories, whereas others are unstable and require higher resolution or need more time to settle into a stationary trajectory. All trajectories which could not be classified automatically as circular helices were tested manually on whether the classification failed because the resolution was insufficient.

The third simulation compares inertial and overdamped trajectories at $T = \SI{0}{\kelvin}$ and demonstrates the convergence of inertial trajectories to the overdamped case. The spherical particle that was simulated for Fig.\ \ref{fig3}(a)-(c) has $\mathcal{H} = \mathrm{I}$ and $\mathcal{M} = \mathrm{I}$, where $\mathrm{I}$ means the $6\times6$--dimensional identity matrix. 
The random particle that was simulated for Fig.\ \ref{fig3}(d)-(f) has
\begin{equation}
    \mathcal{H}_\mathrm{rand} =
    \left(
    \begin{array}{rrr|rrr}
       2.83 &     0.10 &    -1.19 &    -0.16 &     3.84 &    -0.05 \\
       0.10 &     2.43 &    -1.68 &    -2.97 &     0.73 &     0.88 \\
      -1.19 &    -1.68 &     3.60 &     2.13 &    -2.73 &    -0.57 \\ \hline
      -0.16 &    -2.97 &     2.13 &     5.45 &    -1.27 &    -1.05 \\
       3.84 &     0.73 &    -2.73 &    -1.27 &     7.10 &     0.05 \\
      -0.05 &     0.88 &    -0.57 &    -1.05 &     0.05 &     1.66
    \end{array} \right)
\end{equation}
and
\begin{equation}
    \mathcal{M}_\mathrm{rand} =
    \left(
    \begin{array}{rrr|rrr}
       1.19 &     0    &     0    &     0    &     0    &     0    \\
       0    &     1.19 &     0    &     0    &     0    &     0    \\
       0    &     0    &     1.19 &     0    &     0    &     0    \\ \hline
       0    &     0    &     0    &     4.76 &     0.69 &     0.77 \\
       0    &     0    &     0    &     0.69 &     4.88 &    -1.35 \\
       0    &     0    &     0    &     0.77 &    -1.35 &     4.42
    \end{array} \right).
\end{equation}
The principal values of these two matrices were generated in the interval $[1,10]$. The particle's propulsion is $\mathbf{K}_0 = (0,0,3,1,1,1)^\mathrm{T} \cdot 10^{-3}$.
Both the spherical and the randomly shaped particle in Fig.\ \ref{fig3} were simulated for $\SI{20}{\milli\second}$ with $N = 10^6$ time steps.

\section{\label{sec:hydrodynamic_friction_tensor}Reconstruction of the hydrodynamic resistance matrix}
In the overdamped case, it is possible as shown in Ref.\ \cite{KraftWtHEPL2013} to calculate the hydrodynamic resistance matrix from the cross-correlation function of position and orientation as follows.

We define the translational vector $\mathfrak{X}(t) = (\Delta \mathbf{x}(t), \Delta  \mathbf{e}(t))^\mathrm{T}$, composed of the translation vectors of position $\Delta \mathbf{x}(t) = \mathbf{x}(t) - \mathbf{x}(0)$ and orientation $\Delta \mathbf{e}(t) = \frac{1}{2} \smash{\sum_{i=1}^3} \mathbf{e}^{(i)}(t) \times  \mathbf{e}^{(i)}(0)$, where $ \mathbf{e}^{(i)}$ is the $i$th column vector of the orientation matrix $\mathrm{R}$.
The corresponding auto-correlation matrix is
\begin{equation}
	\mathcal{X}(t) = \langle \mathfrak{X}(t) \otimes \mathfrak{X}(t) \rangle \,,
\end{equation}
where the brackets $\langle \cdot \rangle$ denote the ensemble average and $\otimes$ is the outer product.
The short-time limit of the auto-correlation matrix allows calculating the diffusion tensor through
\begin{equation}
  \mathcal{D} = \frac{1}{2} \lim_{t \to 0} \frac{\mathrm{d} \mathcal{X}(t)}{\mathrm{d}t} \,.
  \label{eq:diffusion_tensor_from_correlation}
\end{equation}
We then obtain $\mathcal{H}$ by inverting Eq.\ \eqref{eq:diffusion_tensor_definition}, which relates $\mathcal{D}$ to $\mathcal{H}$, as
\begin{equation}
  \mathcal{H} = \frac{1}{\beta \eta} \mathcal{R} \mathcal{D}^{-1} \mathcal{R}^{-1} \,.
  \label{eq:resistancematrix_tensor_from_correlation}
\end{equation}

In experiments and simulations, the data are discrete. It is therefore useful to discretize Eq.\ \eqref{eq:diffusion_tensor_from_correlation} with step size $\Delta t$:
\begin{equation}
  \mathcal{D} = \frac{1}{2} \frac{
  	\langle \mathfrak{X}(\Delta t) \otimes \mathfrak{X}(\Delta t) \rangle
  	- \langle  \mathfrak{X}(0) \otimes \mathfrak{X}(0) \rangle}{\Delta t}\,.
\end{equation}
Since $\mathfrak{X}(0) = \mathbf{0}$, as can be seen from the definition of $\mathfrak{X}$, we have 
\begin{equation}
  \mathcal{D} = \frac{1}{2} \frac{
  	\langle \mathfrak{X}(\Delta t) \otimes \mathfrak{X}(\Delta t) \rangle}{\Delta t}\,.
  \label{eq:diffusion_tensor_from_correlation_discrete}
\end{equation}
The ensemble average $\langle \cdot \rangle$ is equivalent to the time average. One can therefore either measure many identical particles at the same time or the trajectory of a single particle at many instants in time to obtain the auto-correlation matrix. The latter option is presumably easier in experiments as it requires measuring just one particle.

We find a similar formula for the general case with inertia. Let 
\begin{equation}
\begin{split}
	\mathfrak{V}(t) &= \bigg(\mathbf{v}(t) - \mathbf{v}(0), \\
	&\qquad \frac{1}{2}\sum_{i=1}^3 \mathbf{e}^{(i)}(0) \times \big((\boldsymbol{\omega}(t) - \boldsymbol{\omega}(0)) \times \mathbf{e}^{(i)}(0) \big) \bigg)
\label{eq:H_from_correlation_overdamped}
\end{split}
\end{equation}
be the difference in translational velocity and angular velocity from $t_0 = 0$ on and
\begin{equation}
	\mathcal{V}(t) = \langle \mathfrak{V}(t) \otimes \mathfrak{V}(t)\rangle
\end{equation}
its auto-correlation matrix. Then $\mathcal{H}$ can be calculated from the short-time limit through
\begin{equation}
	\mathcal{H} = \frac{\beta}{\eta} \mathcal{R}_0 \mathcal{M} \bigg(\frac{1}{2} \lim_{t \to 0} \frac{\mathrm{d}\mathcal{V}(t)}{\mathrm{d}t} \bigg) \mathcal{M} \mathcal{R}_0^{-1}\,.
\label{eq:H_from_correlation_inertial}
\end{equation}
This formula is our third main result.

We will now derive this formula. Starting out from Langevin equation \eqref{eq:langevin},
\begin{equation}
  \mathcal{M}  \frac{\partial \mathfrak{v}}{\partial t}
  = [\text{drift terms}] + \mathcal{R}^{-1} \mathbf{k}\,,
\end{equation}
we integrate over a small time step $\Delta t$ and substitute $\mathfrak{V}(\Delta t) = \mathfrak{v}(\Delta t) - \mathfrak{v}(0)$ to obtain
\begin{equation} \label{eq:Langevin_shorttime_integral}
    \mathcal{M} \mathfrak{V}(\Delta t) = 
        [\text{drift terms}] \Delta t 
    + \mathcal{R}^{-1}(0) \sigma \mathbf{W}_{\!\Delta t}\,,
\end{equation} 
which is exact in the limit $\Delta t \to 0$. The vector $\mathbf{W}_{\!\Delta t}$ is a Wiener process, which follows a normal distribution, $\mathbf{W}_{\!\Delta t} \sim N(0, \Delta t)$.
Next, we form the auto-correlation of Eq.\ \eqref{eq:Langevin_shorttime_integral}, neglecting terms of higher order than $\Delta t$, and recall that $\mathcal{M}$ and $\sigma$ are symmetric to get
\begin{equation}
\begin{split}
    \mathcal{M} \big\langle \mathfrak{V}(\Delta t) \otimes \mathfrak{V}(\Delta t) \big\rangle \mathcal{M}
    = 
    \mathcal{R}_0^{-1} \sigma \big\langle \mathbf{W}_{\!\Delta t} \otimes \mathbf{W}_{\!\Delta t} \big\rangle \sigma \mathcal{R}_0\,.
\end{split}
\end{equation}
After inserting $\sigma$ and using that $\big\langle \mathbf{W}_{\!\Delta t} \otimes \mathbf{W}_{\!\Delta t} \big\rangle$ equals the covariance of the Wiener process, which is in general $\mathrm{Cov}(\mathbf{W}_s,\mathbf{W}_t) = \min(s,t)$, we have
\begin{equation}
    \mathcal{M} \mathcal{V}(\Delta t) \mathcal{M}
    = \mathcal{R}_0^{-1} \frac{2 \eta}{\beta} \mathcal{H} \mathcal{R}_0 \Delta t
\,.\end{equation}
Finally, rearranging and taking the limit $\Delta t \to 0$ yields formula \eqref{eq:H_from_correlation_inertial}.

\begin{figure*}[tbhp]
    \centering
    \makebox[\textwidth][c]{\includegraphics[width=\linewidth]{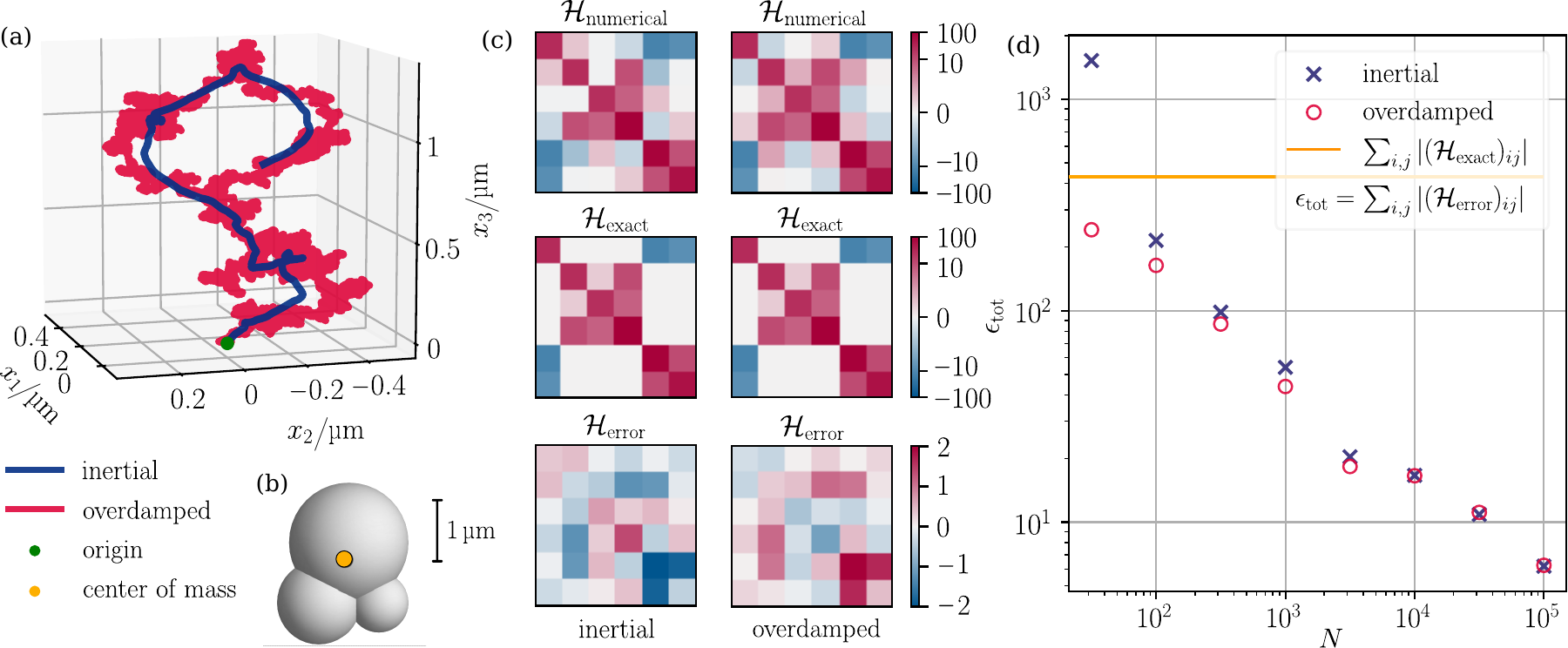}}
    \caption{\label{fig1}The hydrodynamic resistance matrix of a particle can be recovered from its trajectory through either Eq.\ \eqref{eq:H_from_correlation_inertial} or Eq.\  \eqref{eq:resistancematrix_tensor_from_correlation}. (a) A simulated inertial trajectory that takes into account the particle's mass and moment of inertia and a simulated overdamped trajectory are shown for $N = 10^4$ time steps and temperature $T = 300\,\mathrm{K}$. (b) The simulated particle.
    Its hydrodynamic resistance matrix and its moment of inertia tensor can be found in \cref{sec:app:numerical_hydrodynamic_friction_tensor}. (c) Plots that compare the numerical solution $\mathcal{H}_\mathrm{numerical}$ for the reconstructed hydrodynamic resistance matrices with the exact matrix $\mathcal{H}_\mathrm{exact}$ used for the simulations, showing the difference $\mathcal{H}_\mathrm{error}=\mathcal{H}_\mathrm{numerical}-\mathcal{H}_\mathrm{exact}$ for the inertial and overdamped case. The units of all $\mathcal{H}$ are \textmu m, \textmu m$^2$, and \textmu m$^3$.  A similar numerical solution was performed for several numbers of time steps $N$. (d) The total error $\epsilon_\mathrm{tot} = \sum_{i,j} |(\mathcal{H}_\mathrm{error})_{ij}|$ decreases according to a power law for increasing $N$. The inertial formula \eqref{eq:H_from_correlation_inertial} and the overdamped formula \eqref{eq:resistancematrix_tensor_from_correlation} perform equally well for their respective trajectory.}
\end{figure*}

Tests of the overdamped and inertial schemes with trajectories that originate from a simulation based on solving the Langevin equation show that both methods allow to recover $\mathcal{H}$ from the particle's trajectory within their respective regime, as shown in Fig.\ \ref{fig1} for an irregular trimer. The values of all $\mathcal{H}$ are listed in \cref{sec:app:numerical_hydrodynamic_friction_tensor}. The total numerical error is similar for both schemes and diminishes according to a power law when the number of time steps $N$ increases.

Remarkably, the error matrix of the overdamped scheme is close to the negative of the inertial error matrix. This convergence behavior seems to imply that the under- and overdamped schemes approach the exact hydrodynamic resistance matrix as one-sided limits from opposite sides.

In the overdamped regime, the inertial formula does not hold. Similarly, the overdamped formula does not apply in the inertial regime. The new inertial formula is thus no generalization of the overdamped formula. Rather, the two formulas complement each other.

On a side note, the two trajectories in Fig.\ \ref{fig1}(a) for the inertial and overdamped Langevin equation demonstrate the influence of inertia on a particle's random path. The overdamped trajectory is self-similar if we increase the number of steps to zoom in on a part of the trajectory. Inertia breaks this similarity and smooths the particle's path. An inertial particle tends to continue along its way, even when affected by random fluctuations.

\section{\label{sec:simulation_and_analytical_solution}Simulation and analytical solution}
\begin{figure}[tbhp]
    \centering\includegraphics[width=0.96\linewidth]{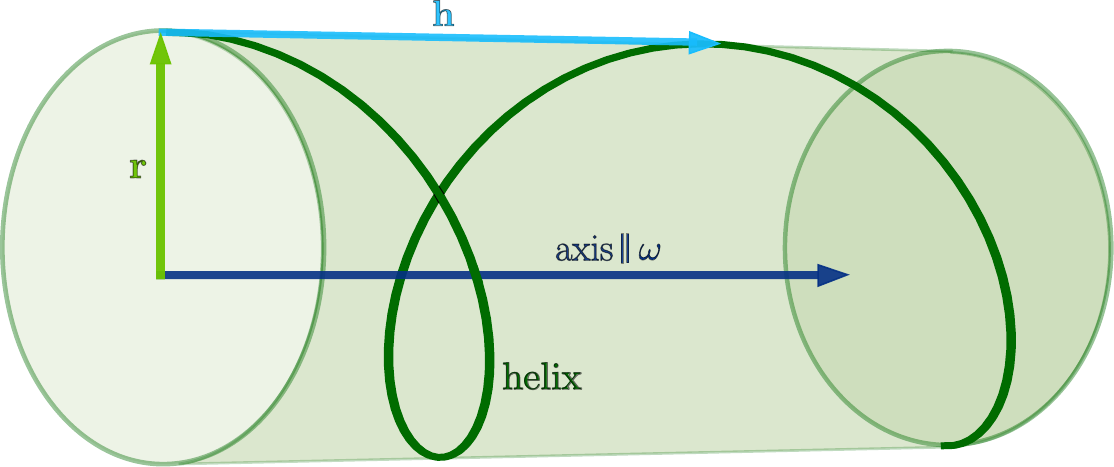}
    \caption{\label{fig2}The right-handed circular helix with radius $\mathbf{r}$ and pitch $\mathbf{h}$ revolves around an axis that points in the direction of the particle's angular velocity $\boldsymbol{\omega}$.}
\end{figure}

The overdamped trajectories of orthotropic self-propelled particles at temperature $T=\SI{0}{\kelvin}$  are circular helices of the form
\begin{equation}
    \mathbf{x}(t) = \mathbf{x}_0 + \mathbf{r} + \mathbf{r} \times \frac{\boldsymbol{\omega}}{\norm{\boldsymbol{\omega}}} \sin(\norm{\boldsymbol{\omega}} t) - \mathbf{r} \cos(\norm{\boldsymbol{\omega}} t) + \frac{\mathbf{h}}{2\pi} \norm{\boldsymbol{\omega}} t
\label{eq:analytical_helix_position}
\end{equation}
with radius
\begin{equation}
    \mathbf{r} = \frac{\boldsymbol{\omega} \times \mathbf{v}_0}{\norm{\boldsymbol{\omega}}^2}\,,
\label{eq:analytical_helix_radius}
\end{equation}
pitch
\begin{equation}
    \mathbf{h} =  2 \pi \frac{\boldsymbol{\omega} \cdot \mathbf{v}_0}{\norm{\boldsymbol{\omega}}^2} \frac{\boldsymbol{\omega}}{\norm{\boldsymbol{\omega}}}\,,
\label{eq:analytical_helix_pitch}
\end{equation}
and the Euclidean norm $\norm{\cdot}$, as was found in Ref.\ \cite{WittkowskiL2012}. See Fig.\ \ref{fig2} for an exemplary illustration of a circular helix.

\begin{figure*}[tbhp]
    \centering\includegraphics[width=\linewidth]{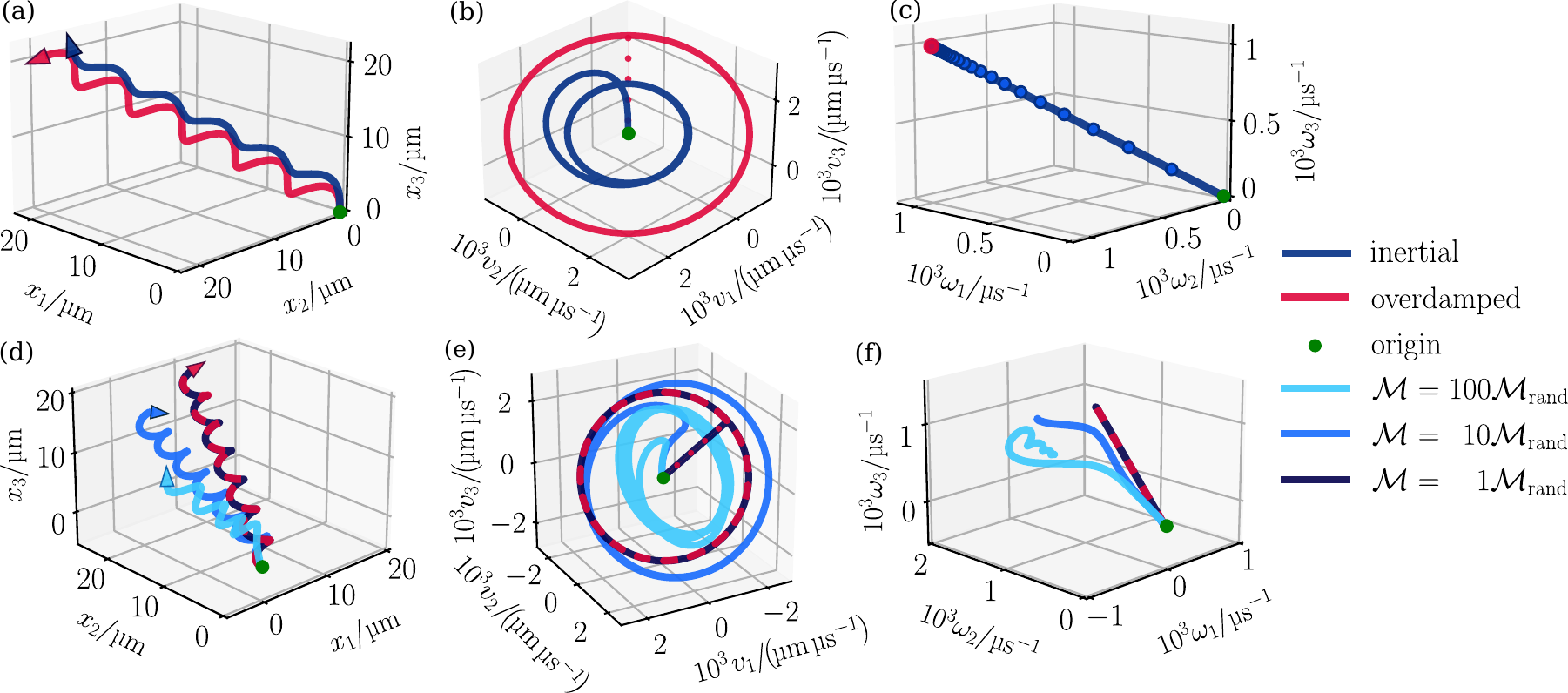}
    \caption{\label{fig3} (a)-(c) Comparison of a spherical particle's inertial trajectory and overdamped trajectory for temperature $T=0\,\mathrm{K}$. See section \ref{sec:simulation_parameters} for the simulation parameters. (a) The trajectories show that the particle moves along a circular helix in position space with an initial transition phase in the inertial case. (b) The particle's translational velocity instantaneously starts to move along a circle in the overdamped case and continuously settles into a circular path in the inertial case. (c) Similarly, the overdamped angular velocity jumps discontinuously to a constant value, whereas the inertial angular velocity exhibits an approximately exponential transition phase.
    (d)-(f) Inertial and overdamped trajectories of an asymmetric particle for three different generalized masses $\mathcal{M}$ at $T=0\,\mathrm{K}$. The inertial trajectories converge to the overdamped one for decreasing masses. Conversely, the initial transition phase becomes more pronounced when the mass increases.}
\end{figure*}
The initial position $\mathbf{x}_0 = \mathbf{x}(0)$ is known from the simulation parameters, but $\mathbf{v}_0$ and $\boldsymbol{\omega}$ do not coincide with whichever initial state $(\mathbf{v}(0), \boldsymbol{\omega}(0))^\mathrm{T}$ was chosen for simulation. In fact, in the overdamped case, the initial generalized velocity $\mathfrak{v}_0$ has no impact on the trajectory since $\mathfrak{v}_1$ is determined independently of $\mathfrak{v}_0$. The particle rather jumps immediately to a stationary trajectory, namely the helix.
Solving the Langevin equation with the knowledge that the trajectory is indeed a circular helix boils down to the question of how $\mathbf{v}_0$ and $\boldsymbol{\omega}$ depend on the experimental boundary conditions or simulation parameters, i.e., properties of the solvent, shape of the particle, and strength of the self-propulsion.

The solution in Ref.\ \cite{WittkowskiL2012} is obtained directly from the overdamped Langevin equation \eqref{langevin_overdamped} with the external generalized force set to zero. Because of $\nu(T=\SI{0}{\kelvin}) = 0$, the drift term and noise are also zero. The translational and rotational parts of the equation then separate into
\begin{equation}
    \mathbf{v} = \mathrm{R}^{-1} \mathrm{K}^{-1} \frac{1}{\eta} \mathbf{F}_0\,, \quad
    \boldsymbol{\omega}  = \mathrm{R}^{-1} \Omega_\mathrm{S}^{-1} \frac{1}{\eta} \mathbf{T}_0\,,
\end{equation}
since $\mathcal{H}$ can be diagonalized for orthotropic particles, and consequently, one obtains
\begin{equation}
    \mathbf{v}_0 = \mathrm{R}_0^{-1} \mathrm{K}^{-1} \frac{1}{\eta} \mathbf{F}_0\,, \quad
    \boldsymbol{\omega}  = \mathrm{R}_0^{-1} \Omega_\mathrm{S}^{-1} \frac{1}{\eta} \mathbf{T}_0
\label{eq:analytical_helix_velocities_overdamped}
\end{equation}
with initial orientation $\mathrm{R}_0 = \mathrm{R}(0)$. The relations \eqref{eq:analytical_helix_velocities_overdamped} determine the helix completely in its orientation, radius, and pitch.

In the inertial case, we find a circular helix after an initial transition phase as the noise-free trajectory for $T=\SI{0}{\kelvin}$. Remarkably, this helix is found regardless of the particle's shape.
This observation is our fourth and final main result.

Figure \ref{fig3}(a) illustrates the difference between the inertial and the overdamped trajectory of a spherical particle. An inertial transition phase at the beginning offsets the inertial trajectory. Radius and pitch differ between the two trajectories.

How can we be sure whether the trajectory is truly a helix rather than a more general spiral? There are two ways to address this question.
The first is to simulate the trajectory so far that the transition phase is passed and to then note the state $(\mathfrak{x}_n, \mathfrak{v}_n)^\mathrm{T}$. If the trajectory is indeed a circular helix, the helix \eqref{eq:analytical_helix_position} determined by $\mathbf{x}_0 = \mathbf{x}_n$, $\mathbf{v}_0 = \mathbf{v}_n$, and $\boldsymbol{\omega} = \boldsymbol{\omega}_n$ should align with the simulated trajectory from time step $n$ onward.
The second, more elegant way is to consider the generalized velocity $\mathfrak{v} = (\mathbf{v}, \boldsymbol{\omega})^\mathrm{T}$ instead of the position. If the translational velocity is circular in velocity space and the angular velocity is constant, the particle moves along a circular helix.

This second approach is exemplified in Fig.\ \ref{fig3}(b) and \ref{fig3}(c) for the spherical particle. We see that inertia leads to a continuous transition between the initial state and the stationary trajectory, whereas the overdamped approximation jumps discontinuously to the stationary trajectory.

Figures \ref{fig3}(d)-(f) show that for decreasing generalized mass $\mathcal{M}$ the inertial trajectory converges to the overdamped trajectory as expected. For increasing $\mathcal{M}$, on the other hand, the transition phase becomes more pronounced while radius, pitch, and orientation of the circular helix change.
These findings are in line with the transition phase found in Ref.\ \cite{ScholzJLL2018} for spherical particles in two spatial dimensions.

To test whether the stationary trajectory is indeed universal, i.e., occurs for every shape, mass, moment of inertia, and internal force and torque of the particle, we evaluated $1000$ simulated inertial trajectories that correspond to different, randomly chosen parameter values. Indeed, we found that all of these trajectories converged to a circular helix.

The regular form of the inertial equilibrium trajectory suggests that, similar to Eqs.\ \eqref{eq:analytical_helix_position}-\eqref{eq:analytical_helix_pitch} for an orthotropic particle in the overdamped case, it is possible to derive an analytical representation of the inertial equilibrium trajectory. However, the irregular inertial transition phase constitutes a serious obstacle for attempts to derive a full analytical solution of the Langevin equation \eqref{eq:langevin} in the noise-free case ($T=\SI{0}{\kelvin}$). It should also be possible to generalize Eqs.\ \eqref{eq:analytical_helix_position}-\eqref{eq:analytical_helix_pitch} towards overdamped nonorthotropic particles. Despite significant efforts, we did not find such analytical expressions. Therefore, their derivations remain a task for future research.

\section{\label{sec:conclusion}Conclusions}
Our central result is the inertial Langevin equation for a self-propelled asymmetric particle. We derived therefrom the corresponding inertial Fokker-Planck equation. This generalized model provides a basis for further research on inertial effects of active Brownian motion.
Moreover, a formula was derived to determine the hydrodynamic resistance matrix from a particle's trajectory. Our new inertial formula was verified numerically and might prove useful for experiments on self-propelled particles where inertia cannot be neglected. Its overdamped counterpart has been tested empirically with artificial colloidal particles in Ref.\ \cite{KraftWtHEPL2013}.
In particular, our simulations have shown that without an external force, the noise-free trajectory of an arbitrarily shaped self-propelled particle converges in time to a circular helix. 
This result is important because it explains on a purely physical basis why many microorganisms move along a helical path.

Qualitative accounts of helical motion of microorganisms date back to 1901 \cite{Jennings1901}. One of the first quantitative analyses is given in Ref.\ \cite{Ludwig1929}. The origin of this helical motion remained, however, subject of speculation: Jennings suggests in Ref.\ \cite{Jennings1901} that the organisms veer to one side because of an asymmetric body shape while rotating around their own axis to achieve a net forward motion in the form of a helix. This behavior would allow the microswimmer to stabilize its track, similar to rifling a bullet. 
Crenshaw argues that the organisms swim along a helix to orient themselves in response to stimuli, like nutrients or light \cite{Crenshaw1996}.
Purcell attributes helical motion to cyclic deformations of an asymmetric body \cite{Purcell1977}.
We have shown in this work that helical motion does not require some specific body shape or mechanism for propulsion. Helices are rather the universal paths for any microswimmer regardless of shape and propulsion. This applies even to large swimmers that experience inertial effects.

So far, only the overdamped circular helix for orthotropic particles has been described analytically. It might be that solutions exist as well for overdamped nonorthotropic particles, or inertial orthotropic and nonorthotropic particles. These solutions might be obtained either through fits of the numerical solutions or by solving the respective Langevin equation analytically.

A natural question to explore in the future is whether the noise-averaged inertial trajectory for $T>0$ is a generalized concho-spiral, as was found for the overdamped case in Ref.\ \cite{WittkowskiL2012}, except for an inertial transition phase.
Furthermore, one could investigate the influence of an external force and torque on the inertial trajectory and compare the results with the categorization of trajectories in the overdamped case from Ref.\ \cite{WittkowskiL2012}. It may be useful to relax the assumption of white noise to colored noise.
One could also extend our model to non-Gaussian diffusion \cite{ChechkinSMS2017}. 
The work presented here focuses on the motion of a single particle. An extension to systems of multiple interacting particles is nonetheless straightforward.

Inertial effects are subject to current research: Vibrobots, for example, exhibit collective behavior near boundaries due to inertia \cite{DebalisBGDVLBBK2018} and self-organize in separate phases \cite{ScholzEP2018}.
Inertia has also been shown to affect the diffusion as well as the swim pressure and Reynolds pressure of active Brownian particles, which could serve to explain phase transitions in active matter \cite{Sandoval2020}.
Another interesting recent development are field theories of inertial active matter that have been employed, for example, to investigate sound propagation and analogues of quantum-tunneling \cite{teVrugtJW2021, teVrugtFHHTW2022}.
Our new description of active Brownian motion may serve to uncover novel inertial effects that occur in collectives of asymmetric self-propelled particles.

\section*{Acknowledgments}
We thank Cornelia Denz for helpful discussions.

\section*{Funding}
J.M.M.\ thanks the Studienstiftung des deutschen Volkes for financial support. 
This work is funded by the Deutsche Forschungsgemeinschaft (DFG, German Research Foundation) -- Project-ID 433682494 -- SFB 1459.

\section*{Competing interests}
There are no conflicts of interest to declare.

\section*{Data availability}
The code for the numerical calculations that were performed for this article is available as Supplementary Material \cite{SI}.

\appendix
\section{\label{sec:app:nondimen_interpretation_time_scale}Interpretation of the characteristic time scale} 
We aim to provide an interpretation of the characteristic time scale $t_\mathrm{c} = m_\mathrm{c}/(\eta l_\mathrm{c})$ defined in Eq.\ \eqref{eq:nondimen_timescale}.
For simplicity, we assume an orthotropic particle in two spatial dimensions on which only hydrodynamic friction acts. The translational and rotational part of the Langevin equation decouple for orthotropic particles, as seen in section \ref{sec:langevin_in_2D}. We need only consider the translational Langevin equation \eqref{eq:langevin_translational_2D},
\begin{equation}
	m \, \partial_t \begin{pmatrix} v_1 \\ v_2 \end{pmatrix}
	= - \eta
	\mathrm{\widetilde{R}}(\phi)
	\begin{pmatrix}
		H_1 & 0 \\
		0   & H_3
	\end{pmatrix}
    \mathrm{\widetilde{R}}^{-1}(\phi)
 \begin{pmatrix} v_1 \\ v_2 \end{pmatrix}.
\end{equation}

Furthermore, we assume that the particle is spherical with radius $r = l_\mathrm{c} / (6 \pi)$ and mass $m = m_\mathrm{c}$. This implies that $H_1 = H_3 = l_\mathrm{c}$, as dictated by the Stokes-Einstein equation, $D = 1 / (6 \pi \eta \beta r)$, for the particle's diffusion coefficient $D$ (which corresponds to the diffusion tensor $\mathcal{D}$ of a general biaxial particle) \cite{Einstein1905c}. We thus obtain
\begin{equation}
	m_\mathrm{c} \, \partial_t \begin{pmatrix} v_1 \\ v_2 \end{pmatrix}
	= - \eta l_\mathrm{c}
 	\begin{pmatrix} v_1 \\ v_2 \end{pmatrix},
\end{equation}
which has the same solution for both $v_1$ and $v_2$ so that only the differential equation
\begin{equation}
	\partial_t v = - \eta \frac{l_\mathrm{c}}{m_\mathrm{c}} v = - \frac{v}{t_\mathrm{c}}
\end{equation}
needs to be solved. Its solution is the exponential function
\begin{equation}
	v = v_0 \mathrm{e}^{- t / t_\mathrm{c}}\,,
\end{equation}
which leads us to the insight that for $t = t_\mathrm{c}$ the velocity has slowed down to $v = v_0 / \mathrm{e}$.

\section{\label{sec:app:motivation_of_rotation}Motivation of the discretized rotation}
Here, we expand on the motivation of Eq.\ \eqref{eq:discrete_rotation} for updating the orientation in each step, without claiming full mathematical rigor, and demonstrate the connection to a vector-based approach.

The rotation matrices $\mathrm{R}$ in three dimensions are elements of the Lie group $\mathrm{SO}(3)$. Its Lie algebra $\mathfrak{so}(3)$ consists of all skew-symmetric matrices. The \mbox{tensor $\mathrm{W}$} generates a rotation through the exponential map
\begin{equation}
	\mathrm{R} = \exp(\mathrm{W} t)\,.
\end{equation}
This suggests that the matrix
\begin{equation}
	\mathrm{A} = \exp(\mathrm{W} \mathrm{d}t)
\end{equation}
is an infinitesimal rotation. For sufficiently small $\Delta t$, the Taylor expansion of $\mathrm{A}$ around $\mathrm{I}$ is
\begin{equation}
\begin{split}
	\mathrm{A} &= \mathrm{I} + \mathrm{W} \Delta t + \frac{1}{2} (\mathrm{W} \Delta t)^2  + \dotsb \\
	&\approx  \mathrm{I} + \mathrm{W} \Delta t\,.
\end{split}
\end{equation}
A particle with orientation $\mathrm{R}_n$ that rotates with angular velocity determined by $\mathrm{W}_n$ would be expected to perform the minute rotation $\mathrm{A}_{n}$ within $\Delta t$ so that the linear operator $\mathrm{A}_{n}$ applied to $\mathrm{R}_n$ gives Eq.\ \eqref{eq:discrete_rotation}. For an in-depth discussion of angular momentum and rotations, see Ref.\ \cite{Rose1995}.

Similarly, one can calculate $\mathrm{R}_{n+1}$ within vector calculus by interpreting the columns of the rotation matrices $\mathrm{R}$ as vectors $\mathbf{e}^{(i)}$ for $i = 1,2,3$ through
\begin{equation}
  \frac{\mathrm{d} \mathbf{e}^{(i)}}{\mathrm{d}t}
  = \boldsymbol{\omega} \times \mathbf{e}^{(i)} = \mathrm{W} \, \mathbf{e}^{(i)}\,,
\label{eq:discrete_rotation_vector_continuous}
\end{equation}
just as in the classical mechanical relation $\mathrm{d}\mathbf{x}/\mathrm{d}t = \boldsymbol{\omega} \times \mathbf{x}$ for the rotation of a vector $\mathbf{x}$ with angular velocity $\boldsymbol{\omega}$.
Equation \eqref{eq:discrete_rotation_vector_continuous} can be discretized as
\begin{equation}
	  \mathbf{e}^{(i)}_{n+1} = \mathbf{e}^{(i)}_n
	  + \Delta t \, \boldsymbol{\omega} \times \mathbf{e}^{(i)}_n\,.
\label{eq:discrete_rotation_vector_discrete}
\end{equation}
The approaches per matrices \eqref{eq:discrete_rotation} and vectors \eqref{eq:discrete_rotation_vector_discrete} are computationally roughly equally expensive.

\begin{widetext}
\section{\label{sec:app:overdamped_drift_term_calculation}Explicit calculation of the overdamped drift term}
A tedious, yet functional way to calculate $\nabla_\mathrm{R} \mathrm{R}^\mathrm{T} \mathrm{A} \mathrm{R}$ is to write the rotation matrix $\mathrm{R}$ in terms of Euler angles $\boldsymbol{\varpi} = (\phi, \theta, \chi)^\mathrm{T}$ with definitions \eqref{eq:elementary_rotation_matrices_definition_2} and \eqref{eq:elementary_rotation_matrices_definition_3} of the elementary rotation matrices: 
\begin{equation}
\begin{split}
    \mathrm{R}(\boldsymbol{\varpi}) = 
    \begin{pmatrix*}[r]
    \cos{(\phi)} \cos{(\theta)} \cos{(\chi)} - \sin{(\phi)} \sin{(\chi)} &
    \sin{(\phi)} \cos{(\theta)} \cos{(\chi)} + \cos{(\phi)} \sin{(\chi)} &
    - \sin{(\theta)} \cos{(\chi)} \\
    - \cos{(\phi)} \cos{(\theta)} \sin{(\chi)} - \sin{(\phi)} \cos{(\chi)} &
    - \sin{(\phi)} \cos{(\theta)} \sin{(\chi)}
    +\cos{(\phi)} \cos{(\chi)} &
     \sin{(\theta)} \cos{(\chi)} \\
    \cos{(\phi)} \sin{(\theta)} & \sin{(\phi)} \sin{(\theta)} & \cos{(\theta)}
    \end{pmatrix*}.
\end{split}
\end{equation}
We apply the rotational gradient operator $\nabla_\mathrm{R} = \mathrm{i} \mathbf{L}$, which can be written with Eq.\ \eqref{eq:fokker-planck_derivation_iL_euler} as 
\begin{equation}
	\mathrm{i} \mathbf{L} =
	\begin{pmatrix}
	    - \cos(\phi) \cot(\theta) \, \partial_\phi
	    - \sin(\phi) \, \partial_\theta
	    + \cos(\phi) \csc(\theta) \, \partial_\chi \\
	    - \sin(\phi) \cot(\theta) \, \partial_\phi
	    + \cos(\phi) \, \partial_\theta
	    + \sin(\phi) \csc(\theta) \, \partial_\chi \\
	    \partial_\phi
	\end{pmatrix} ,
\end{equation}
to the expression $\mathrm{R}^\mathrm{T}\mathrm{A}\mathrm{R}$. It is then possible to identify elements of $\mathrm{R}$ in the result. This yields
\begin{equation}
\begin{split}
	\nabla_\mathrm{R} \mathrm{R}^\mathrm{T} \mathrm{A} \mathrm{R}
	&= \begin{pmatrix}
	(A_{12} - A_{21}) R_{31} - (A_{13} - A_{31})  R_{21} + (A_{23} - A_{32}) R_{11} \\	
	(A_{12} - A_{21}) R_{32} - (A_{13} - A_{31}) R_{22} + (A_{23} - A_{32}) R_{12} \\
	(A_{12} - A_{21}) R_{33} - (A_{13} - A_{31}) R_{23} + (A_{23} - A_{32}) R_{13}
	\end{pmatrix} \\
	&= (A_{23} - A_{32}, - (A_{13} - A_{31}), A_{12} - A_{21}) \, \mathrm{R} \\
	&= (\mathrm{R}^\mathrm{T} \Delta \mathrm{A})^\mathrm{T}\,,
\end{split}
\end{equation}
where
$\Delta \mathrm{A} =
(A_{23} - A_{32}, - (A_{13} - A_{31}), A_{12} - A_{21})^\mathrm{T}$ is a vector of the differences between the off-diagonal elements of matrix $\mathrm{A}$.

\section{\label{sec:app:numerical_hydrodynamic_friction_tensor}Numerical results for the hydrodynamic resistance matrix}

The trajectories that exemplify the convergence of Eqs.\ \eqref{eq:H_from_correlation_overdamped} and \eqref{eq:H_from_correlation_inertial} were simulated with the exact hydrodynamic resistance matrix of a trimer (three differently sized spheres glued together) obtained in Ref.\ \cite{KraftWtHEPL2013} through hydrodynamic calculations:
\begin{equation}
    \mathcal{H}_\mathrm{exact}
    =
    \left(
    \begin{array}{rrr|rrr}
         27.9 &  0.0 &  0.0 &   0.0 & -12.6 & -7.2 \\
          0.0 & 26.1 &  0.3 &  11.0 &   0.0 &  0.0 \\
          0.0 &  0.3 & 24.8 &   6.0 &   0.0 &  0.0 \\ \hline
          0.0 & 11.0 &  6.0 & 104.4 &   0.0 &  0.0 \\
        -12.6 &  0.0 &  0.0 &   0.0 &  90.2 & 11.2 \\
         -7.2 &  0.0 &  0.0 &   0.0 &  11.2 & 58.9
    \end{array} \right) .
    \label{eq:H_trimer_exact}
\end{equation}
The overdamped formula \eqref{eq:discrete_langevin_overdamped} applied to the overdamped trajectory yields
\begin{equation}
    \mathcal{H}_\mathrm{num} =
    \left(
    \begin{array}{rrr|rrr}
         27.67 &    -0.34 &     0.02 &     0.26 &   -12.58 &    -6.98 \\
         -0.34 &    26.39 &     0.68 &    12.03 &     0.95 &     0.11 \\
          0.02 &     0.68 &    24.27 &     5.78 &    -0.35 &    -0.06 \\ \hline
          0.26 &    12.03 &     5.78 &   103.46 &     0.41 &    -0.30 \\
        -12.58 &     0.95 &    -0.35 &     0.41 &    92.04 &    12.73 \\
         -6.98 &     0.11 &    -0.06 &    -0.30 &    12.73 &    59.29
    \end{array} \right) ,
\end{equation}
at $10^4$ steps with absolute error
\begin{equation}
    \mathcal{H}_\mathrm{err} =
    \left(
    \begin{array}{rrr|rrr}
          -0.23 &    -0.34 &     0.02 &     0.26 &     0.02 &     0.22 \\
          -0.34 &     0.29 &     0.38 &     1.03 &     0.95 &     0.11 \\
           0.02 &     0.38 &    -0.53 &    -0.22 &    -0.35 &    -0.06 \\ \hline
           0.26 &     1.03 &    -0.22 &    -0.94 &     0.41 &    -0.30 \\
           0.02 &     0.95 &    -0.35 &     0.41 &     1.84 &     1.53 \\
           0.22 &     0.11 &    -0.06 &    -0.30 &     1.53 &     0.39
    \end{array} \right) ,
\end{equation}
whereas the inertial formula \eqref{eq:discrete_langevin} applied to the inertial trajectory gives
\begin{equation}
    \mathcal{H}_\mathrm{num} =
    \left(
    \begin{array}{rrr|rrr}
          28.19 &     0.36 &    -0.03 &    -0.28 &   -12.64 &    -7.45 \\
           0.36 &    25.85 &    -0.09 &    10.00 &    -0.94 &    -0.10 \\
          -0.03 &    -0.09 &    25.37 &     6.25 &     0.39 &     0.06 \\ \hline
          -0.28 &    10.00 &     6.25 &   105.67 &    -0.34 &     0.32 \\
         -12.64 &    -0.94 &     0.39 &    -0.34 &    88.63 &     9.76 \\
          -7.45 &    -0.10 &     0.06 &     0.32 &     9.76 &    58.60
    \end{array} \right) ,
\end{equation}
at $10^4$ steps with absolute error
\begin{equation}
    \mathcal{H}_\mathrm{err} =
    \left(
    \begin{array}{rrr|rrr}
       0.29 &     0.36 &    -0.03 &    -0.28 &    -0.04 &    -0.25 \\
       0.36 &    -0.25 &    -0.39 &    -1.00 &    -0.94 &    -0.10 \\
      -0.03 &    -0.39 &     0.57 &     0.25 &     0.39 &     0.06 \\ \hline
      -0.28 &    -1.00 &     0.25 &     1.27 &    -0.34 &     0.32 \\
      -0.04 &    -0.94 &     0.39 &    -0.34 &    -1.57 &    -1.44 \\
      -0.25 &    -0.10 &     0.06 &     0.32 &    -1.44 &    -0.30
    \end{array} \right) .
\end{equation}
All values are stated in \textmu m for the upper left, \textmu m$^2$ for the upper right and lower left, and \textmu m$^3$ for the lower right block of each matrix.

To calculate the moment of inertia, the irregular trimer is modeled as three spheres of radius $0.7\,\si{\micro\meter}$, $0.5\,\si{\micro\meter}$, and $1\,\si{\micro\meter}$ with relative coordinates $(-0.5, 0, 0)^\mathrm{T}\,\si{\micro\meter}$, $(0.5, 0, 0)^\mathrm{T}\,\si{\micro\meter}$, and $(0, 1, 0)^\mathrm{T}\,\si{\micro\meter}$, respectively \cite{KraftWtHEPL2013}.
We choose as reference point for the moment of inertia the center of mass $(-0.07, 0.72, 0)^\mathrm{T}\,\si{\micro\meter}$.
The moment of inertia is then
\begin{equation}
    \mathrm{J} =
    \left(
    \begin{array}{rrr}
         3.22 & -0.03 &     0 \\
        -0.03 &  2.36 &     0 \\
            0 &     0 &  3.63
    \end{array} \right) \,\si{\pico\gram \,\micro\meter^2} \,.
\end{equation}
\end{widetext}

\bibliography{refs}
\end{document}